\documentclass[prb,aps,amsmath,amssymb,showpacs,twocolumn]{revtex4-1}

\usepackage{amssymb,graphicx,float,dcolumn,bm,subfigure,color}

\begin{document}

\title{Quantum Hall Effect of Two-Component Bosons at Fractional and Integral Fillings}

\author{Ying-Hai Wu and Jainendra K. Jain}
\affiliation{Department of Physics, The Pennsylvania State University, University Park, PA 16802}

\date{\today}

\begin{abstract}
We investigate the feasibility of many candidate quantum Hall states for two-component bosons in the lowest Landau level. We identify interactions for which spin-singlet incompressible states occur at filling factors $\nu=2/3$, $4/5$ and $4/3$, and spin-partially-polarized states at filling factors $3/4$ and $3/2$, where ``spin" serves as a generic label for the two components. We study ground states, excitations, edge states and entanglement spectrum for systems with up to $16$ bosons, and construct explicit trial wave functions to clarify the underlying physics. The composite fermion theory very accurately describes the ground states as well as excitations at $\nu=2/3$, $4/5$ and $3/4$, although it is less satisfactory for the $\nu=3/2$ state. For $\nu=4/3$ a ``non-Abelian spin-singlet" state, which is the exact ground state of a 3-body contact interaction, has been proposed to occur even for a 2-body contact interaction; our trial wave functions are very accurate for the excitations of the 3-body interaction, but they do not describe the excitations of the 2-body interaction very well. Instead, we find that the $\nu=4/3$ state is more likely to be a spin-singlet state of reverse-flux-attached composite fermions at filling $\nu^*=4$. We also consider incompressible states at integral filling factors $\nu=1$ and $2$. The incompressible state at $\nu=1$ is shown to be well described by the parton-based Jain spin-singlet wave function, and the incompressible state at $\nu=2$ as the spin-singlet state of reverse-flux-attached composite fermions at $\nu^*=2$, which provides an example of the bosonic integer topological phase.
\end{abstract}

\maketitle

\section{Introduction}
\label{intro}

The study of two-component fractional quantum Hall (FQH) effect has revealed a tremendous amount of new physics. The earlier studies were performed on GaAs systems~\cite{Eisenstein,Duspin,Yeh,Kukushkin,Yusa01,Gros07,Hayakawa,Melinte,Smet01,Kraus,Smet,Tiemann}, where the Land\'e g-factor is small and therefore both components of spin can be active at relatively small magnetic fields. More recently, two-component FQH effect has been studied in systems where valleys play the role of spin, as in AlAs quantum wells~\cite{Shayegan,Gokmen} and H-terminated Si(111) surface~\cite{Kane12}; here the Zeeman energy is large enough to freeze the spin degree of freedom for typical experimental parameters. In graphene, the two components could be either spins or valleys, depending on parameters~\cite{Kim,Yacoby}. Experiments have shown that in general, FQH states with several spin/valley polarization can occur at a given filling factor, and transitions between them can be caused by tuning the Zeeman/valley splitting. These level crossing transitions are understood in terms a competition between the composite fermion (CF) cyclotron energy and the Zeeman/valley splitting. A quantitative understanding of this physics has been achieved through Halperin's multi-component wave functions~\cite{Halperin}, and more generally through the theory of spinful composite fermions~\cite{Wu93,Park98,Davenport1,Jain}.

Given a rich diversity of strongly correlated states of fermions involving the spin physics, it is natural to ask what new physics can be learned from the study of two-component Bose gases, such as those made up of two hyperfine spin states of the same atoms, in the FQH regime. Neutral bosons can in principle be driven into the FQH regime by rapid rotation~\cite{Review}. Strongly correlation among particles is achieved as the number of vortices $N_V$ in a rotating Bose-Einstein condensate (BEC) becomes comparable with the number of atoms $N$ as quantified by the filling factor $\nu=N/N_V$. For simplicity, we will refer to the two components as spins, but the results apply to any two-component bosons for which the interaction is (approximately) independent of the component index. There has been much recent study of bosonic quantum Hall states~\cite{Cooper1,Cooper2,Cooper3,Viefers,Korslund,Wilkin,Manninen,Sinova,Schweikhard,Chang,Bargi,Papenbrock,Christensson,Borgh,Regnault1,Regnault2}. It has been shown that the vortex lattice that forms at large $\nu$ melts and that a series of FQH states appear at various filling factors, which include, for appropriately chosen interactions, Laughlin~\cite{Laughlin}, Jain~\cite{Comp}, Moore-Read~\cite{Pfaffian} and Read-Rezayi~\cite{Read} states. While FQH effect in cold atom systems has not yet been observed in a convincing manner, substantial progress in that direction has been reported~\cite{Gemelke}. Other ingenious methods to simulate the effect of magnetic field have also been explored~\cite{Lin,Peter,Dalibard}. 

We consider below two-component bosons in the FQH regime. Aside from the experimental interest, a motivation for studying this problem is the possibility of realization of new structures that are not available in electronic FQH effect. In particular, we will see that some FQH states require a tuning of the interaction, which is more easily accomplished in ultracold atomic systems.

The theoretical study of FQH effect has relied on the notion of formation of emergent quasiparticles, description in terms of which provides a tangible way to understand the physical properties of an inherently hard quantum $N$-body problem. The physics of emergent quasiparticles is captured by appropriate wave functions, which, in turn, represent topological phases. To fully classify all topological phases is a formidable task, but progress has been made in the context of topological insulators and superconductors~\cite{Schnyder,Kitaev}. However, one can take specific examples and ask if they occur for models with realistic interactions. We consider in this article several bosonic spin-singlet and spin-partially-polarized states, and ask for what kinds of interaction they would be realized. Some of these support excitations with Abelian braid statistics, whereas some with non-Abelian braid statistics. 

The plan of this paper is as follows. In Sec.~\ref{wave} we introduce all of the trial wave functions that we study in the subsequent sections. Sec.~\ref{model} describes our model, and our methods for evaluating the wave functions, exact diagonalization and entanglement spectra. Sec.~\ref{fractional} presents the results for bosons at fractional fillings and Sec.~\ref{integral} for integral fillings. Sec.~\ref{conclusion} summarizes the conclusions of our study.

\section{Trial Wave Functions}
\label{wave}

In this section, we shall use the symmetric gauge on disk geometry where the lowest Landau level (LLL) wave functions are particular simple as given by
\begin{eqnarray}
\phi_m (z) = \frac{z^m\exp\left(-|z|^2/4\right)}{\sqrt{2\pi 2^m m!}}
\end{eqnarray}
where $z=x+iy$ is the complex coordinate of particles on the disk. The ubiqutous exponential factor will be omitted in the rest of this paper. A bosonic Fock state is represented using symmetric monomials and a many-body state is the superposition of all monomials with appropriate quantum numbers. We use the convention that the coordinates $\{z^\uparrow\}$ and $\{z^\downarrow\}$ denote, respectively, spin-up and spin-down particles, whereas $\{z\}$ denote {\em all} particles.

The general wave function of two-component bosons (with $N_\uparrow$ spin-up and $N_\downarrow$ spin-down bosons) at filling factor $\nu$ has the form
\begin{equation}
\chi_{\nu} = \mathcal{S} \left[ \Psi_{\nu}(\{z\}) u_1 \cdots u_{N_\uparrow} d_{1} \cdots d_{N_\downarrow} \right]
\end{equation}
where $\Psi_{\nu}(\{z\}$ is the spatial part, $u$ and $d$ refer to the two components, and $\mathcal{S}$ denotes symmerization. It is sufficient to consider $\Psi_{\nu}(\{z\})$ provided it satisfies appropriate symmetries. An acceptable wave function with spin $S=S_z$ must satisfy Fock's cyclic condition, which means that the state $\chi_{\nu}$ is annihilated by an attempt to antisymmetrize a spin-down particle with respect to the spin-up particles. This condition is satisfied for the wave functions considered below.

(i) The Halperin 221 state at $\nu=2/3$ state is given by
\begin{eqnarray}
\Psi^{221}_{\frac{2}{3}} (\{z\}) = \prod_{i<j}(z^\uparrow_i-z^\uparrow_j)^2 (z^\downarrow_i-z^\downarrow_j)^2 \prod_{i,j}(z^\uparrow_i-z^\downarrow_j) 
\label{Halperin221}
\end{eqnarray}
This form of multi-component wave functions were introduced by Halperin for electronic FQH states~\cite{Halperin}. The Halperin 221 wave function vanishes as the third power of distance between particles when two particles are brought together, {\em regardless} of their spin configuration. It is the exact ground state for the contact interaction $\sum_{i<j} \delta(z_i-z_j)$. 

(ii) The Jain's CF (JCF) states at $\nu=n/(n{\pm}1)$ are given by
\begin{widetext}
\begin{eqnarray}
\Psi^{[n_\uparrow,n_\downarrow]}_{\frac{n}{n+ 1}}(\{z\}) &=& {\cal P}_{\rm LLL} \biggl[ \Phi_{n_\uparrow}(\{z^\uparrow\}) \Phi_{n_\downarrow}(\{z^\downarrow\}) J(\{z\}) \biggr] 
\label{JainCF1} \\
\Psi^{[-n_\uparrow,-n_\downarrow]}_{\frac{n}{n-1}}(\{z\}) &=& {\cal P}_{\rm LLL} \biggl[ \Phi_{-n_\uparrow}(\{z^\uparrow\}) \Phi_{-n_\downarrow}(\{z^\downarrow\}) J(\{z\}) \biggr] 
\label{JainCF2}
\end{eqnarray}
\end{widetext}
where $\Phi_{-n_\uparrow}\equiv \Phi^*_{n_\uparrow}$, $\Phi_{-n_\downarrow}\equiv \Phi^*_{n_\downarrow}$ and $J(\{z\})=\prod_{i<j}(z_i-z_j)$ is the Jastrow factor for {\em all} particles; $\Phi_{n_\uparrow}$ and $\Phi_{n_\downarrow}$ are two Slater determinants for the spin-up and spin-down particles at fillings $n_\uparrow$ and $n_\downarrow$, respectively, and their complex conjugates $\Phi_{-n_\uparrow}$ and $\Phi_{-n_\downarrow}$ represent filled LL states in opposite magnetic field; $n=n_\uparrow + n_\downarrow$; and the symbol ${\cal P}_{\rm LLL}$ represents the LLL projection operator. The spin polarization is given by 
\begin{equation}
P = \frac{n_\uparrow - n_\downarrow}{n_\uparrow + n_\downarrow}
\end{equation}
Those with $n_\uparrow = n_\downarrow$ are spin-singlet, while those with $n_\uparrow \neq n_\downarrow$ ({\em i.e.} odd $n$) are spin-partially-polarized (or spin-polarized). 

These wave functions are closely related to those studied previously for electronic FQH effect\cite{Wu93,Park98}, where they represent the physics of electrons capturing two vortices to turn into composite fermions, which then form integer quantum Hall (IQH) states. In the present case, the bosons capture one vortex each to form composite fermions, which experience a reduced effective magnetic field $B^*=B-\rho hc/e$ ($B$ is the external field and $\rho$ is the density) and condense into IQH states (with filling factor denoted as $\nu^*$) to produce incompressibility. An intuitive reason for why bosons convert into composite fermions is because this builds good correlations that keep the particles away from one another and thus reduce the interaction energy. For $n_{\downarrow}=0$ these wave functions reduce to fully spin polarized bosons which haven been considered previously~\cite{Regnault1,Chang}. The wave functions in Eq.~(\ref{JainCF1}) and Eq.~(\ref{JainCF2}) are interpreted as the states in which composite fermions fill $n_\uparrow$ spin-up and $n_\downarrow$ spin-down $\Lambda$ levels ($\Lambda$Ls), where $\Lambda$Ls are Landau-like levels of composite fermions. 

A noteworthy aspect of the analogy to IQH effect is that it goes beyond the ground state and also allows construction of wave functions for the excitations of the $\nu=n/(n{\pm}1)$ state in terms of the known excitations of the IQH states. In fact, the CF theory implies a one-to-one correspondence between the excitations at $\nu^*=n$ and those at $\nu$, because an IQH wave function with a given spin and angular momentum quantum numbers produces, through Eqs.~(\ref{JainCF1}) or~(\ref{JainCF2}), a wave function at $\nu$ with the same quantum numbers. In particular, neutral and charged excitations of the IQH state at $\nu^*=n$ produce neutral and charged excitations of the state at $\nu$. In what follows, {\em the JCF wave function $\Psi^{[\pm n_\uparrow,\pm n_\downarrow]}_{n/(n{\pm}1)}$ will collectively represent wave functions for the ground state as well as neutral and charged excitations.}

We study below $\Psi^{[1,1]}_{2/3}$, $\Psi^{[2,2]}_{4/5}$, $\Psi^{[-2,-2]}_{4/3}$, $\Psi^{[2,1]}_{3/4}$, $\Psi^{[-2,-1]}_{3/2}$ and $\Psi^{[-1,-1]}_{2}$, including ground state and excitations. We note that for $n_\uparrow=n_\downarrow=1$ the ground state wave function is given by 
\begin{eqnarray}
\Psi^{[1,1]}_{\frac{2}{3}, {\rm G.S.}} (\{z\}) = {\cal P}_{\rm LLL} \left[ \Phi_{1}(\{z^\uparrow\}) \Phi_1(\{z^\downarrow\}) J(\{z\}) \right] \nonumber
\\ = \prod_{i<j} (z^\uparrow_i-z^\uparrow_j) (z^\downarrow_i-z^\downarrow_j) \prod_{i,j}(z_i-z_j)
\end{eqnarray}
(no LLL projection is required in this case) which is identical to the Halperin 221 wave function. In other words, the Halperin-221 state is interpreted as the $\nu^*=2$ spin-singlet state of composite fermions. This interpretation also allows a construction of the excitations of the 2/3 state by correspondence with the excitations of the $\nu^*=2$ spin-singlet IQH state $\Phi_1(\{z^\uparrow\}) \Phi_1(\{z^\downarrow\})$. 

(iii) The simplest non-Abelian spin-singlet (NASS) state~\cite{Ardonne1,Ardonne2} at filling factor $\nu=2k/3$ can be written as a symmetrized product of $k$ copies of the Halperin $221$ state
\begin{equation}
\Psi^{\rm NASS}_{\frac{2k}{3}} (\{z\}) = {\cal S}_{\uparrow\downarrow} \biggl[ \Psi^{\rm 221}_{\frac{2}{3}} (\{z^\alpha\}) \Psi^{\rm 221}_{\frac{2}{3}} (\{z^\beta\}) \cdots \Psi^{\rm 221}_{\frac{2}{3}} (\{z^k\}) \biggr] 
\label{NASSWave}
\end{equation}
where the particles are divided into $k$ groups with coordinates $\{z^\alpha\},\{z^\beta\},\cdots,\{z^k\}$ and ${\cal S}_{\uparrow\downarrow}$ denotes the {\em separate} symmetrization of the spin-up and spin-down particles. It may be viewed as a spin-singlet generalization of Read-Rezayi $\mathbb{Z}_k$ states~\cite{Read} whose excitations obey non-Abelian braiding statistics~\cite{Ardonne1,Ardonne2}. It is the exact zero energy ground state of a model $(k+1)$-body contact interaction. It has recently been suggested~\cite{Grab,Furukawa} that the 4/3 NASS state may be realized even for the 2-body contact interaction. 

We will also study excitations of this state. The quasihole excitations, obtained by adding flux quanta, also have zero energy for the $(k+1)$-body interaction, and can be explicitly constructed~\cite{Ardonne3,Estienne}. The neutral excitations and the quasiparticles of the $(k+1)$-body Hamiltonian are nontrivial and do not have zero energy. To construct trial wave functions for them, we generalize Eq.~(\ref{NASSWave}) to 
\begin{widetext}
\begin{equation}
\Psi^{\rm NASS}_{\frac{2k}{3}} (\{z\}) = {\cal S}_{\uparrow\downarrow} \biggl[ \Psi^{[1,1]}_{2\over 3} (\{z^\alpha\}) \Psi^{[1,1]}_{2\over 3} (\{z^\beta\}) \cdots \Psi^{[1,1]}_{2\over 3} (\{z^k\}) \biggr] 
\label{NASSWave2}
\end{equation}
\end{widetext}
This reproduces the wave function of Eq.~(\ref{NASSWave}) when all factors $\Psi^{[1,1]}_{2/ 3}$ are chosen as the ground states ({\em i.e.} the Halperin 221 state), but also produces excitations by appropriate choice of excited states on the right hand side. For example, the lowest energy neutral excitations corresponds to a CF exciton in a single factor $\Psi^{[1,1]}_{2/3}$. This approach for constructing excitations follows a ``multipartite CF" representation investigated recently to study the excitations of the Moore-Read state~\cite{Sreejith1,Sreejith2,Rodriguez} and the Read-Rezayi $\mathbb{Z}_3$ state~\cite{Sreejith3}. The NASS state can also be generalized to produce other candidate incompressible states by replacing the Halperin $221$ state with $\Psi^{[\pm n_\uparrow,\pm n_\downarrow]}_{n/(n\pm 1)}$. 

(iv) Moran {\em et al.}~\cite{Moran} recently studied the Jain spin-singlet (JSS) wave function for fermions, which they argued contains topological $d$-wave pairing structure. We consider here its bosonic analog at $\nu=1$
\begin{equation}
\Psi^{\rm JSS}_1 (\{z\}) = {\cal P}_{\rm LLL} \biggl[ \Phi_2(\{z\}) \prod_{i<j}(z^\uparrow_i-z^\uparrow_j) (z^\downarrow_i-z^\downarrow_j)  \biggr]
\label{JSSWave}
\end{equation}
where $\Phi_2$ is the wave function of {\em two} filled Landau levels. This does not belong to the $\Psi^{[\pm n_\uparrow,\pm n_\downarrow]}_{n/(n\pm 1)}$ states considered above, but follows from the parton construction of FQH states~\cite{Parton}. In this construction, each boson is viewed as the bound states of two fictitious species of fermions (partons), one of which carries spin while the other is spinless. The spinful  fermions occupy the spin-singlet state at $\nu=2$ whereas the spinless ones occupy the fully spin-polarized state at $\nu=2$. The fermionic version of this state (obtained by multiplication by another full Jastrow factor) describes a spin-singlet incompressible state at $\nu=1/2$; it was introduced in Ref.~[\onlinecite{Parton}] and considered as a possible candidate for the spin-singlet $5/2$ FQH state~\cite{Belkhir1,Belkhir2}, but was abandoned when it was realized that the Coulomb $5/2$ state is fully spin-polarized. 

We will see below in Sec.~\ref{integral} that this state is realizable for a 2-body interaction. This result is of interest because $\Psi^{\rm JSS}_1$ is the simplest ``parton" state that goes beyond the CF interpretation (all states of composite fermions admit a parton construction but the converse is not true).  The excitations of this state are more complicated. One may naively expect that the low-lying energy levels can be obtained by creating excitations in either $\Phi_2$ or $\prod_{i<j}(z^\uparrow_i-z^\uparrow_j) (z^\downarrow_i-z^\downarrow_j)$ in Eq.~(\ref{JSSWave}). However, it turns out that neither of them gives a very accurate description of the excitations, as we shall see in Sec.~\ref{integral}. 

Many of the above wave functions involve $\Phi_n$, the Slater determinant wave function of $n$ filled LLs, on the right hand side. While $\Phi_n$ is uniquely defined for a {\em compact} geometry, where the number of single-particle states in each Landau level is finite, that is not the case in the disk geometry. For example, in the disk geometry $\Phi_2$ can be defined with $N_1$ particles in the lowest Landau level and $N_2$ particles in the second Landau with the constraints that $N=N_1+N_2$. Different possible choices of $N_1,N_2$ complicates the analysis of the edge excitations of the states involving $\Phi_2$, as has been found to be the case for spin-polarized fermions at 2/5~\cite{Sreejith4,Rodriguez2}.

\section{Models and Methods}
\label{model}

We consider a bosonic system with two internal states in a rapidly rotating harmonic trap. These neutral particles experience forces in the rotating reference frame which mathematically has the same description as charged particles moving in a uniform magnetic field. We specialize to the case where single-particle cyclotron energy is much larger than the many-body gap, so the bosons can be treated as in the lowest Landau level only and effects due to Landau levels mixing are negelected. The number of particles, the number of spin-up particles and the number of spin-down particles are denoted using $N$, $N_{\uparrow}$ and $N_{\downarrow}$, respectively. 

\subsection{Spherical and disk geometry}

We will use the spherical geometry~\cite{Haldane} for most of our calculations. The flux enclosed by the sphere is denoted as $2Q$, which is related to the numbers of particles $N$ and the filling factor $\nu$ via $2Q=N/\nu-{\cal S}_h$. The quantity ${\cal S}_h$ is called the ``shift." Sometimes there is an ambiguity when two states at different fillings ``alias," {\em i.e.}, occur at the same flux. In such cases, it is important to study several values of $N$ to draw unambiguous information. The compact spherical geometry is very convenient for studying the bulk properties of a FQH state, due to absence of edges. For studying the structure of edge excitations, there are two ways of proceeding. One can study either the states in the disk geometry, or the entanglement spectrum in the spherical geometry~\cite{Li} (see Sec.~\ref{model} D). 

The single-particle eigenstates on a sphere are the so-called monopole harmonics~\cite{Yang}
\begin{widetext}
\begin{eqnarray}
Y_{Qlm} &=& N_{Qlm} (-1)^{l-m} u^{Q+m} v^{Q-m} \sum^{l-m}_{s=0} (-1)^s \binom{l-Q}{s} \binom{l+Q}{l-m-s} (u^*u)^s (v^*v)^{l-Q-s}
\end{eqnarray}
where $l=Q+n$ ($n$ is the Landau level index) is the angular momentum, $m$ is the $z$ component of angular momentum, and $\theta$ and $\phi$ are the azimuthal and radial angles. The spinor coordinates $u=\cos(\theta/2)e^{i\phi/2}$, $v=\sin(\theta/2)e^{-i\phi/2}$ and the normalization coefficient $N_{Qlm}$ is
\begin{eqnarray}
N_{Qlm} &=& \left( \frac{2l+1}{4\pi}\frac{(l-m)!(l+m)!}{(l-Q)!(l+Q)!} \right)^{1/2}
\end{eqnarray}

\subsection{Lowest Landau level projection}

When Eq.~(\ref{JainCF1}) and Eq.~(\ref{JainCF2}) are constructed on a sphere, the flux $2Q^*$ experienced by composite fermions, that is, the flux of the IQH states $\Phi_{\pm n_\uparrow}(\{z^\uparrow\}) \Phi_{\pm n_\downarrow}(\{z^\downarrow\})$, is related to the actual flux by $2Q=2Q^*+(N-1)$. Once the IQH states are constructed using the above single-particle wave functions, we multiply them by the Jastrow factor $J$ and then project the products to the LLL. An efficient Jain-Kamilla method~\cite{Kamilla} has been developed that applies to states of the form ${\cal P}_{\rm LLL}J^{2p} \Phi_n$, where the projected wave function can be constructed for rather large $N$ without the need for expanding it in basis functions. This method requires {\em even} exponent of $J$ for technical reasons. In Ref.~[\onlinecite{Chang}] this method was applied to spinless bosons, by writing ${\cal P}_{\rm LLL}J \Phi_n$ as $J^{-1}{\cal P}_{\rm LLL}J^2 \Phi_n$. Unfortunately, this method does not work for spin-singlet sates, because $J^{-1}{\cal P}_{\rm LLL}J^2 \Phi_{n_\uparrow,n_\downarrow}$ is a singular, non-normalizable wave function, as ${\cal P}_{\rm LLL}J^2 \Phi_{n_\uparrow,n_\downarrow}$ does not vanish when two particles with opposite spins coincide. Therefore, we must evaluate the LLL projection by using its expansion in terms of the symmetric monomials for the spin-singlet states~\cite{Dev92,Wu93}. The following identity of monopole harmonics discovered by Wu and Yang~\cite{Yang} are useful in the LLL projection
\begin{eqnarray}
Y_{Q_1l_1m_1} Y_{Q_2l_2m_2} &=& (-1)^{m_3-Q_3} \sum_{l_3} S(\{Q_i,l_i,m_i\}) Y_{Q_3,l_3,m_3}
\end{eqnarray}
where we have defined the following quantities
\begin{eqnarray}
S(\{Q_i,l_i,m_i\}) &=& (-1)^{l_1+l_2+l_3} \left( \frac{(2l_1+1)(2l_2+1)(2l_3+1)}{4\pi} \right)^{1/2} F^{l_1l_2l_3}_{-m_1-m_2m_3} F^{l_1l_2l_3}_{Q_1Q_2-Q_3} \\
F^{l_1l_2l_3}_{m_1m_2m_3} &=& \frac{(-1)^{l_1-l_2-m_3}}{\sqrt{2l_3+1}} \langle l_1,m_1;l_2,m_2 | l_3,-m_3 \rangle
\end{eqnarray}
Here $Q_3=Q_1+Q_2$, $m_3=m_1+m_2$ and $\langle l_1,m_1;l_2,m_2 | l_3,m_3 \rangle$ is the Clebsch-Gordon coefficient. 

The computational time to perform the LLL projection grows factorially with the number of particles, since one must consider all possible permutations of the indices. As a result, $N=14$ or $16$ is the maximum number of particles that we can study in a reasonable amount of time.
\end{widetext}

\subsection{Exact diagonalization}

Interaction between particles can be parametrized by the Haldane pseudopotential in the 2-body case and their generalizations in the 3-body case~\cite{Simon,Davenport2}. We study Hamiltonians containing 2-body and 3-body interactions, denoted as $H_2$ and $H_3$, respectively:
\begin{eqnarray}
H_2 &=& \sum_\alpha \sum_{ij} c_\alpha \left[ P_{ij}(\alpha,1) + P_{ij}(\alpha,0) \right] \\
H_3 &=& \sum_{ijk} \left[ P_{ijk}(0,3/2) + P_{ijk}(0,1/2) \right]
\end{eqnarray}
where $P_{ij}(L,S)$ projects out a pair of particles $i,j$ with {\em relative} angular momentum $L$ and total spin $S$, and $P_{ijk}(L,S)$ projects out a triple of particles $i,j,k$ with relative angular momentum $L$ and total spin $S$. The natural interaction for bosons is the contact interaction, which corresponds, in units of $c_0$, to
\begin{equation}
H^{\rm con}_2= \sum_{ij}  \left[ P_{ij}(0,1) + P_{ij}(0,0) \right] 
\end{equation}
This will be the interaction used unless otherwise stated. Non-zero values for $c_1$ and $c_2$ in $H_2$, and the 3-body Hamiltonian $H_3$ will also be used sometimes, to stabilize certain interesting states. Since the interaction is rotationally invariant and spin-independent, the energy eigenstates are also eigenstates of orbital angular momentum $\widehat{L}^2$ [with eigenvalue $L(L+1)$] and spin angular momentum $\widehat{S}^2$ [with eigenvalue $S(S+1)$]. In the figures shown below, the energy levels are labeled by their angular momentum and spin quantum numbers $L$ and $S$ and are also shifted horizontally according to their $S$ values for clarity. 

To study edge excitations, we use the disk geometry. The Hamiltonian can also be represented using 2-body Haldane pseudopotentials
\begin{eqnarray}
{\widetilde H}_2 &=& \sum_\alpha \sum_{ij} {\widetilde c}_\alpha \left[ P_{ij}(\alpha,1) + P_{ij}(\alpha,0) \right] 
\nonumber \\ &+& \omega_c ({\widehat L}_z-L_0)
\end{eqnarray}
where ${\widehat L}_z$ is the $z$-component angular momentum operator and the term $\omega_c ({\widehat L}_z-L_0)$ is due to a parabolic confinement potential whose strength is controlled by the parameter $\omega_c$. We choose the coefficients ${\widetilde c}_\alpha$ to have the same values as their counterparts in the spherical geometry Hamiltonian and tune the coefficient $\omega_c$ to make sure that the state at angular momentum $L_0$ has the lowest energy, where the counting of edge excitations starts.

\subsection{Entanglement spectrum}

In addition to comparing the wave functions with exact eigenstates obtained in finite systems, we also study the entanglement spectrum~\cite{Li} in some cases, because it can provide additional insight into the physics of the FQH states. In particular, it has been found that the entanglement spectrum contains information about the edge excitations; specifically, entanglement spectrum can reproduce the counting of the edge states (which provides a method of study edge excitations in the spherical geometry). To obtain the entanglement spectrum for an incompressible ground state $|\Psi\rangle$, one divides the Hilbert space into two parts labeled as $A$ and $B$ and then decomposes the ground state as
\begin{eqnarray}
|\Psi\rangle &=& \sum_{\alpha\beta} C_{\alpha\beta} |\Psi^A_\alpha\rangle \otimes |\Psi^B_\beta\rangle \nonumber \\
&=& \sum_{i} e^{-\xi_i/2} |\Psi^A_i\rangle \otimes |\Psi^B_i\rangle 
\end{eqnarray}
where $|\Psi^A_\alpha\rangle$ and $|\Psi^B_\beta\rangle$ are two sets of basis states for $A$ and $B$, respectively. The second step is achieved through a singular value decomposition (SVD) of the matrix $C_{\alpha\beta}$, which also changes the basis states to $|\Psi^A_i\rangle$ and $|\Psi^B_i\rangle$. A plot of the ``eigenvalues" $\xi_i$ versus the conserved quantum numbers in region $A$ comprises the entanglement spectrum. We shall calculate the ``real space entanglement spectrum"~\cite{Sterdyniak,Dubail,Rodriguez2} (RSES), where the cut is made along the equator and the southern hemisphere is chosen as $A$, with $N^A_{\uparrow}$ ($N^A_{\downarrow}$) spin-up (spin-down) particles. Due to the choice of cut, the levels in the RSES can be labeled by the $z$ component of the total angular momentum $L^A_z$ and the total spin quantum number $S^A$ of the particles in $A$. To compare the edge excitations with the RSES, we will calculate energy spectra on disk geometry when the edge counting cannot be predicted {\em exactly}. For example, the counting of edge excitations of the NASS state can be predicted in several ways and does not require exact diagonalization, but the counting of the edge excitations of JCF state and JSS state are more complicated. 

\begin{figure}
\includegraphics[width=0.45\textwidth]{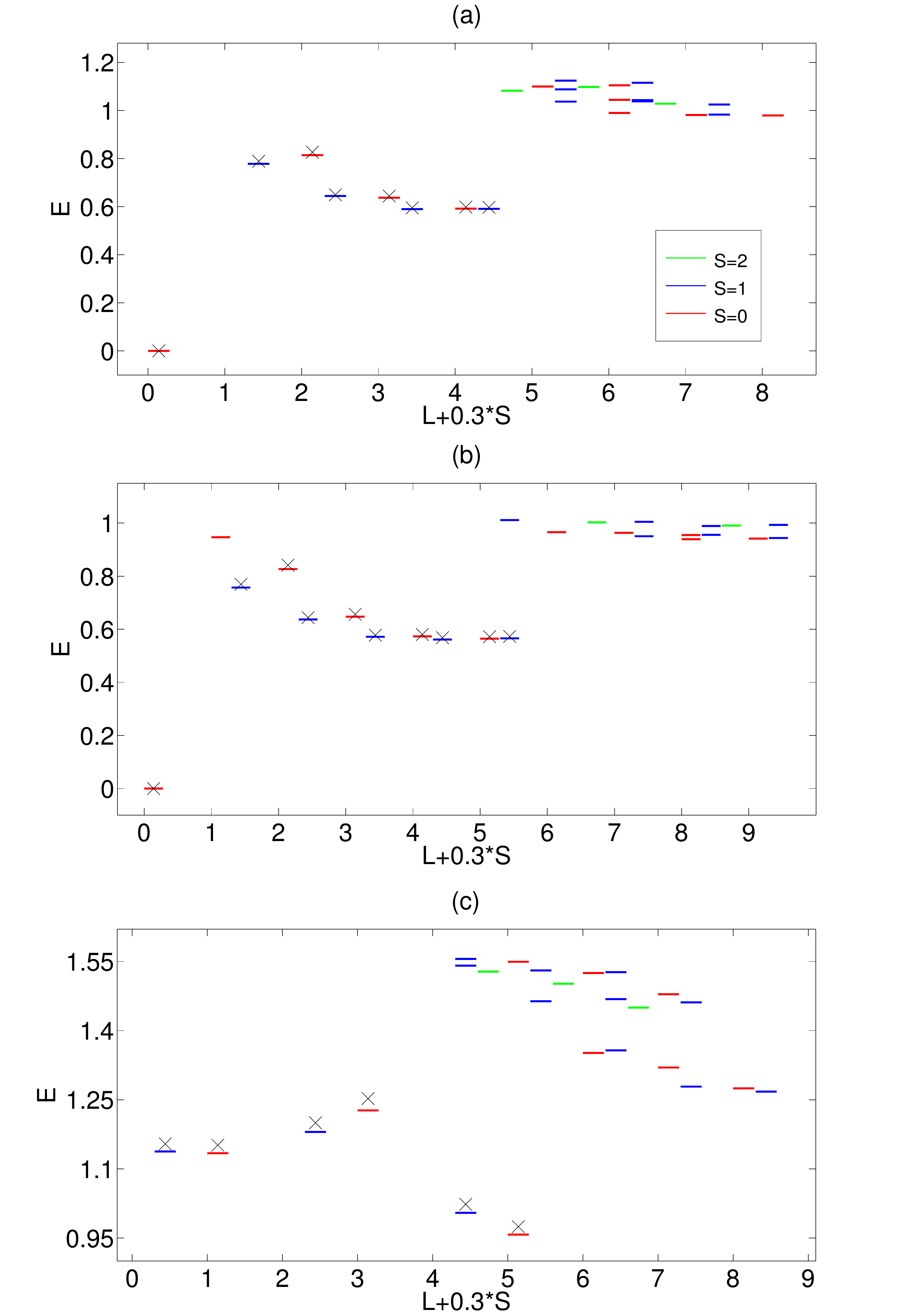}
\caption{Energy spectra (lines) of the $\nu=2/3$ state for the 2-body contact Hamiltonian $H^{\rm con}_2$. The lines are colored according to their spin quantum numbers and are also shifted in the horizontal direction for clarity. The same conventions are used in all other figures. The crosses represent the energies of the wave functions $\Psi^{[1,1]}_{2/3}$ for the ground and excited states. The panels correspond to (a) $N_{\uparrow}=4$, $N_{\downarrow}=4$ and $2Q=10$; (b) $N_{\uparrow}=5$, $N_{\downarrow}=5$ and $2Q=13$; (c) $N_{\uparrow}=5$, $N_{\downarrow}=5$ and $2Q=12$. The inset in (a) shows the color scheme for all panels. Panels (a) and (b) correspond to incompressible states where the uniform ground state has $L=0$ and $S=0$, and the excitations are neutral particle-hole pairs of composite fermions. Panel (c) corresponds to a system containing two quasiparticles; the low energy band contains all possible states of these quasiparticles.}
\label{Figure1}
\end{figure}

\begin{figure}
\includegraphics[width=0.45\textwidth]{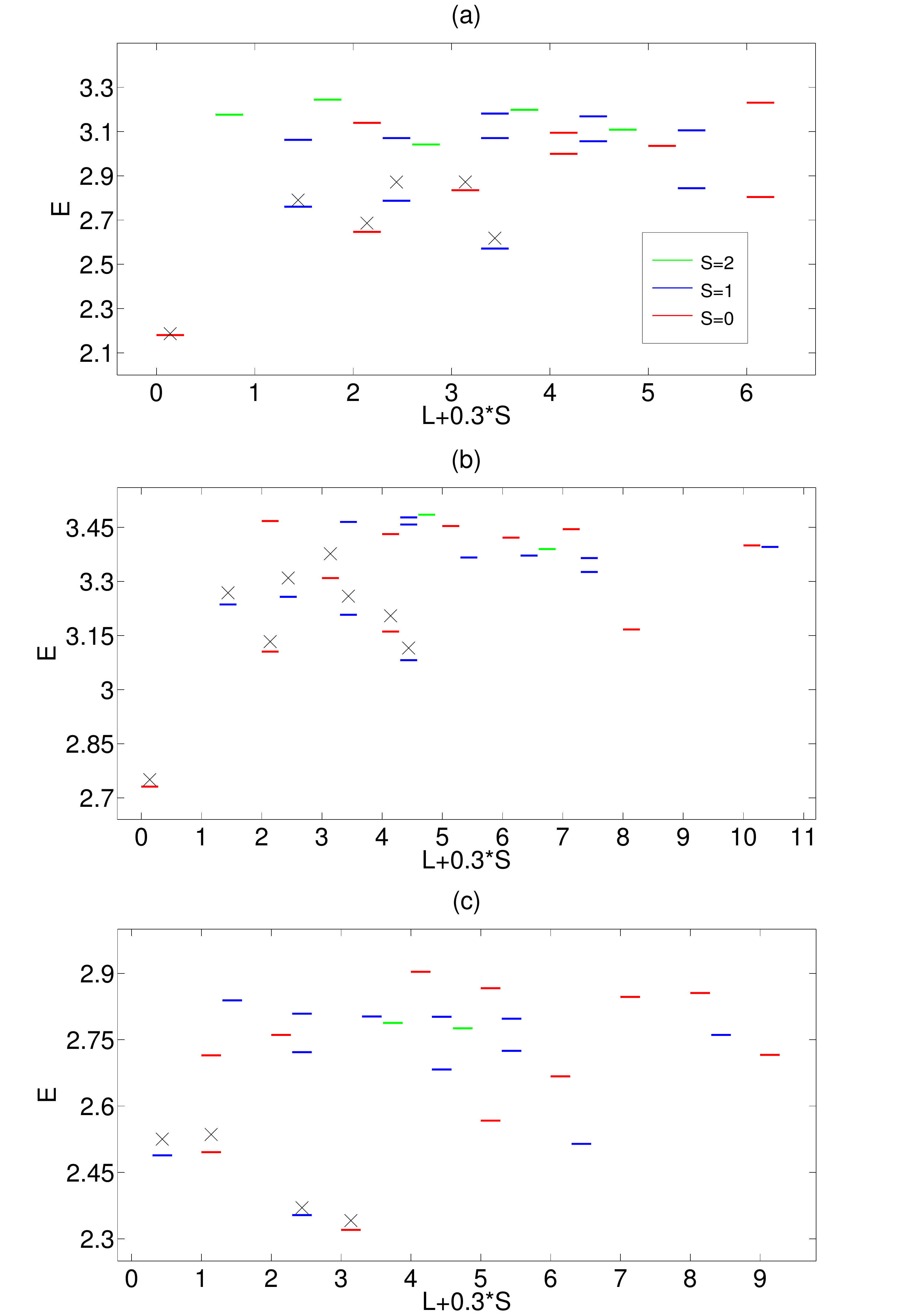}
\caption{Energy spectra of the $\nu=4/5$ state for the Hamiltonian $H^{\rm con}_2$. The crosses represent the energies of the wave functions $\Psi^{[2,2]}_{4/5}$. (a) $N_{\uparrow}=4$, $N_{\downarrow}=4$ and $2Q=7$; (b) $N_{\uparrow}=6$, $N_{\downarrow}=6$ and $2Q=12$; (c) $N_{\uparrow}=5$, $N_{\downarrow}=5$ and $2Q=10$. The inset in (a) shows the color scheme for all panels. Panels (a) and (b) correspond to incompressible states where the uniform ground state has $L=0$ and $S=0$, and the excitations are neutral particle-hole pairs of composite fermions. Panel (c) corresponds to a system containing two quasiholes; the low energy band contains all possible states of these quasiholes.}
\label{Figure2}
\end{figure}

\begin{figure}
\includegraphics[width=0.45\textwidth]{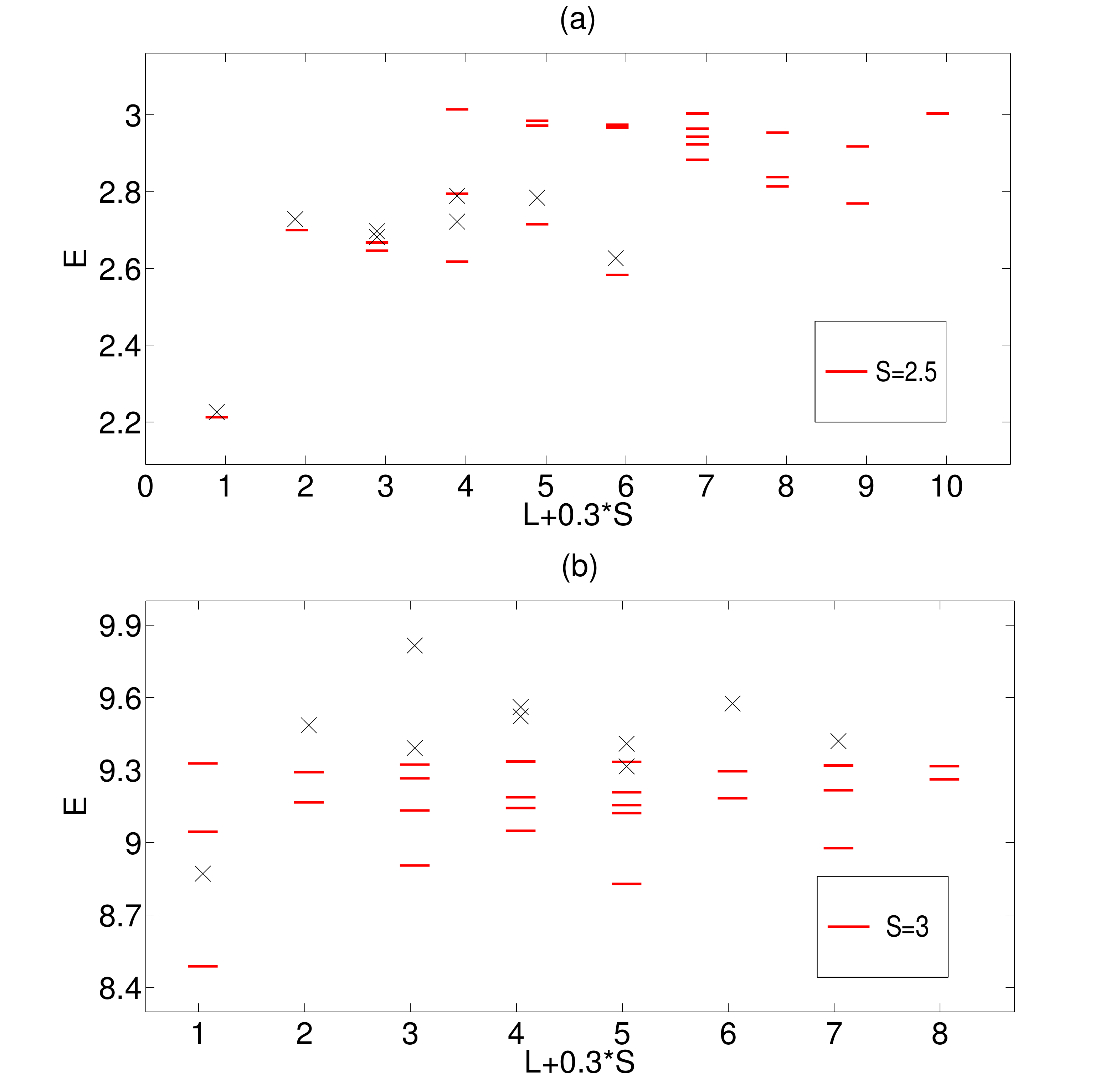}
\caption{(a) Energy spectrum of the $\nu=3/4$ state for the Hamiltonian $H^{\rm con}_2$ with $N_{\uparrow}=3$, $N_{\downarrow}=8$ and $2Q=12$. The crosses represent the energies of the wave functions $\Psi^{[1,2]}_{3/4}$. (b) energy spectrum of the $\nu=3/2$ state for the Hamiltonian $H_2^{\rm con}$ with $N_{\uparrow}=4$, $N_{\downarrow}=10$ and $2Q=10$. The crosses represent the energies of the states $\Psi^{[-1,-2]}_{3/2}$. The insets show the color schemes for the panels.}
\label{Figure3}
\end{figure}

\begin{figure}
\includegraphics[width=0.45\textwidth]{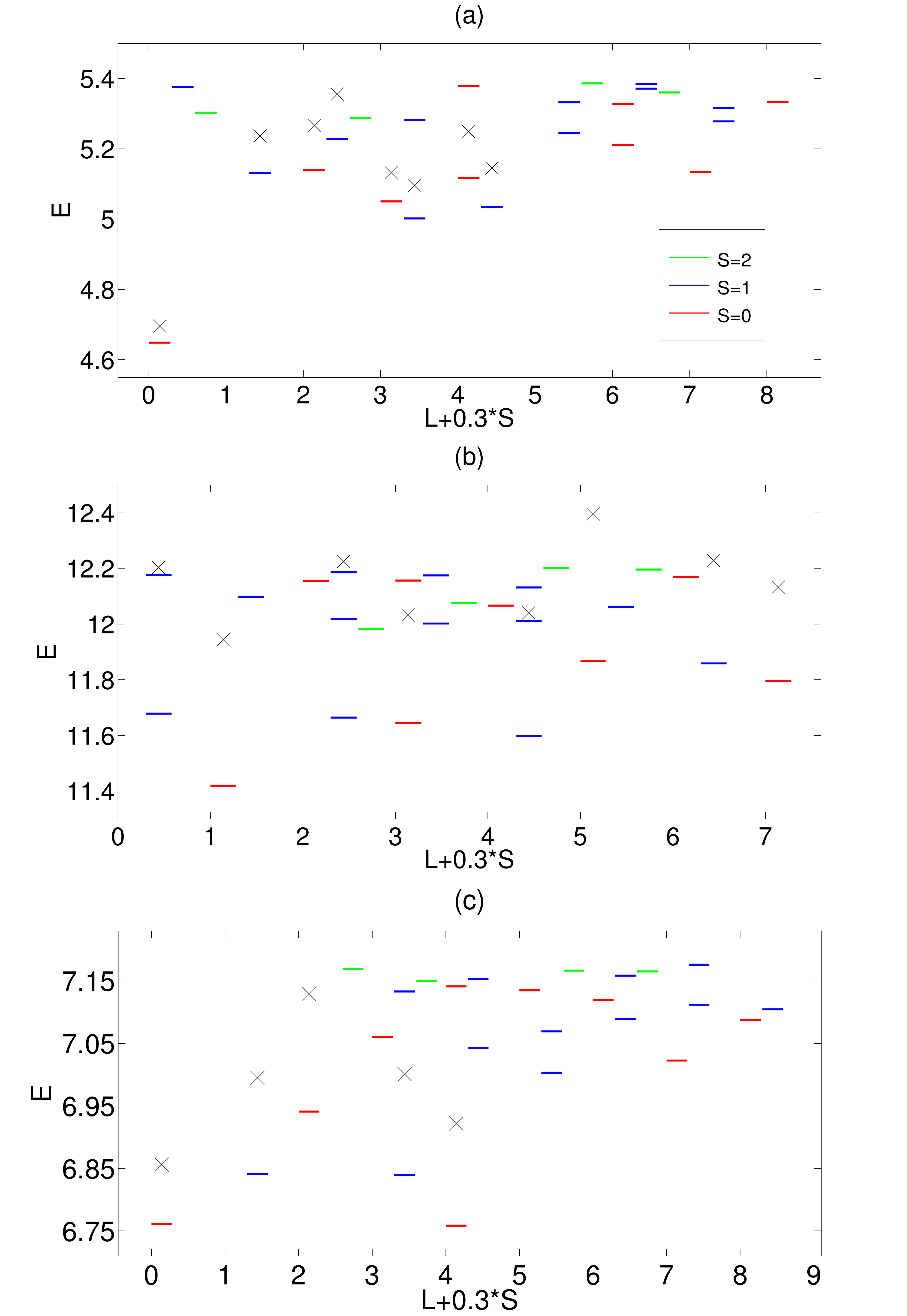}
\caption{Energy spectra of the $\nu=4/3$ state for the Hamiltonian $H^{\rm con}_2$. The crosses represent the energies of the wave functions $\Psi^{[-2,-2]}_{4/3}$. (a) $N_{\uparrow}=6$, $N_{\downarrow}=6$ and $2Q=10$; (b) $N_{\uparrow}=7$, $N_{\downarrow}=7$ and $2Q=8$; (c) $N_{\uparrow}=7$, $N_{\downarrow}=7$ and $2Q=11$. The inset of panel (a) shows the color scheme for all panels.}
\label{Figure4}
\end{figure}

\begin{figure}
\includegraphics[width=0.45\textwidth]{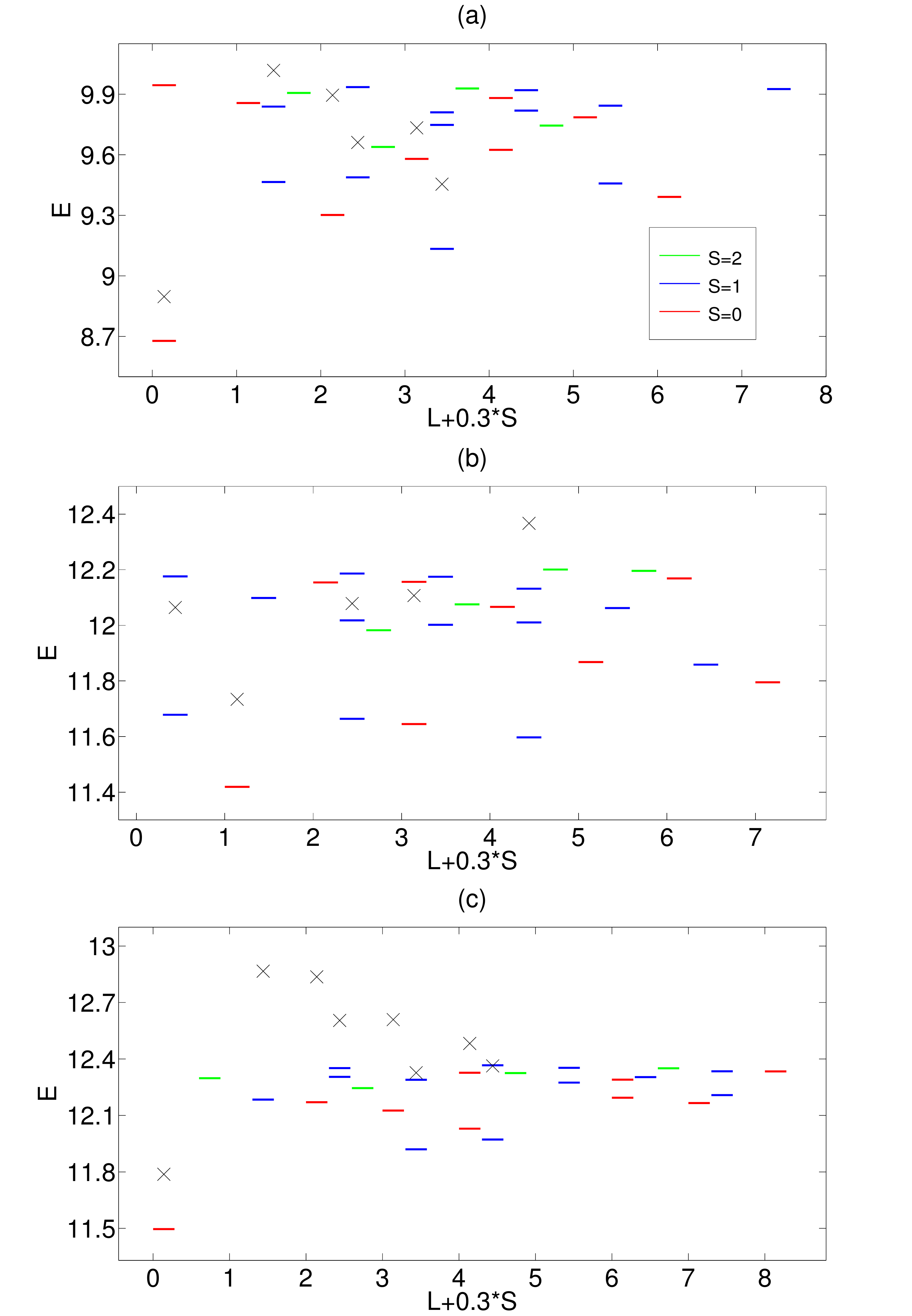}
\caption{Energy spectra of the $\nu=4/3$ state for the Hamiltonian $H^{\rm con}_2$. The crosses represent the energies of the wave functions $\Psi^{\rm NASS}_{4/3}$ obtained from the spinful bipartite CF theory. (a) $N_{\uparrow}=6$, $N_{\downarrow}=6$ and $2Q=7$; (b) $N_{\uparrow}=7$, $N_{\downarrow}=7$ and $2Q=8$; (c) $N_{\uparrow}=8$, $N_{\downarrow}=8$ and $2Q=10$. The inset of panel (a) shows the color scheme for all panels.}
\label{Figure5}
\end{figure}

\begin{figure}
\includegraphics[width=0.45\textwidth]{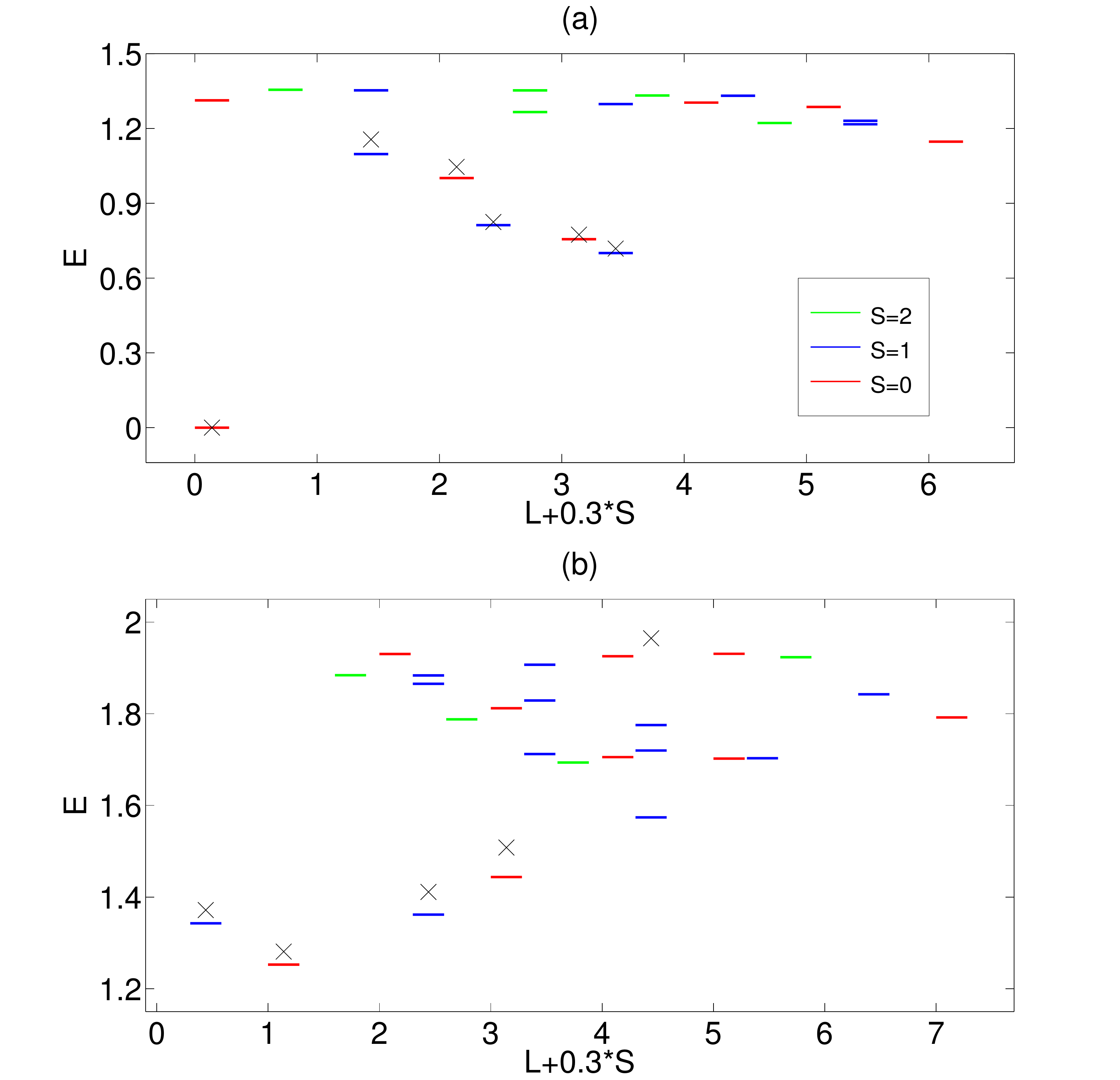}
\caption{Energy spectra of the $\nu=4/3$ state for the 3-body Hamiltonian $H_3$. The crosses represent the energies of the wave functions $\Psi^{\rm NASS}_{4/3}$ obtained from the spinful bipartite CF theory. (a) $N_{\uparrow}=6$, $N_{\downarrow}=6$ and $2Q=7$; (b) $N_{\uparrow}=7$, $N_{\downarrow}=7$ and $2Q=8$. The inset of panel (a) shows the color scheme for both panels.}
\label{Figure6}
\end{figure}

\begin{figure}
\includegraphics[width=0.45\textwidth]{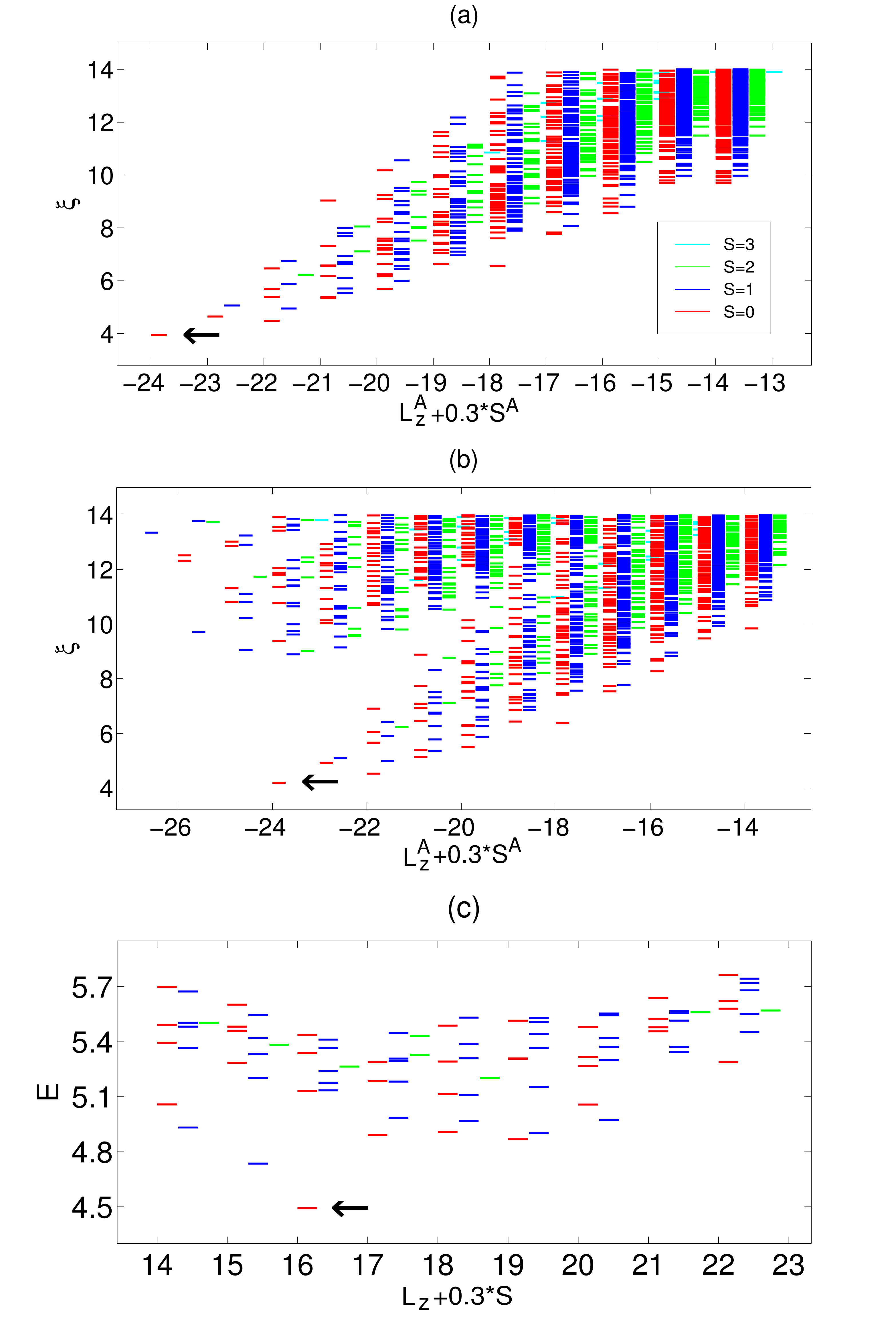}
\caption{RSES and edge excitations of the $\nu=4/3$ NASS state. (a) $N_{\uparrow}=8$, $N_{\downarrow}=8$, $N^A_{\uparrow}=4$, and $N^A_{\downarrow}=4$, using the exact NASS state; (b) $N_{\uparrow}=8$, $N_{\downarrow}=8$, $N^A_{\uparrow}=4$, and $N^A_{\downarrow}=4$, using the ground state of the 2-body Hamiltonian $H_2$; (c) $N_{\uparrow}=4$ and $N_{\downarrow}=4$, energy spectrum on disk geometry of the Hamiltonian ${\widetilde H}_2$ with confinement potential parameter $\omega_c=0.4$. The inset of panel (a) shows the color scheme for all panels. The arrow in panel (c) indicates the ground state and the arrows in panel (a) and panel (b) show the corresponding levels in the RSES.}
\label{Figure7}
\end{figure}

\begin{figure}
\includegraphics[width=0.45\textwidth]{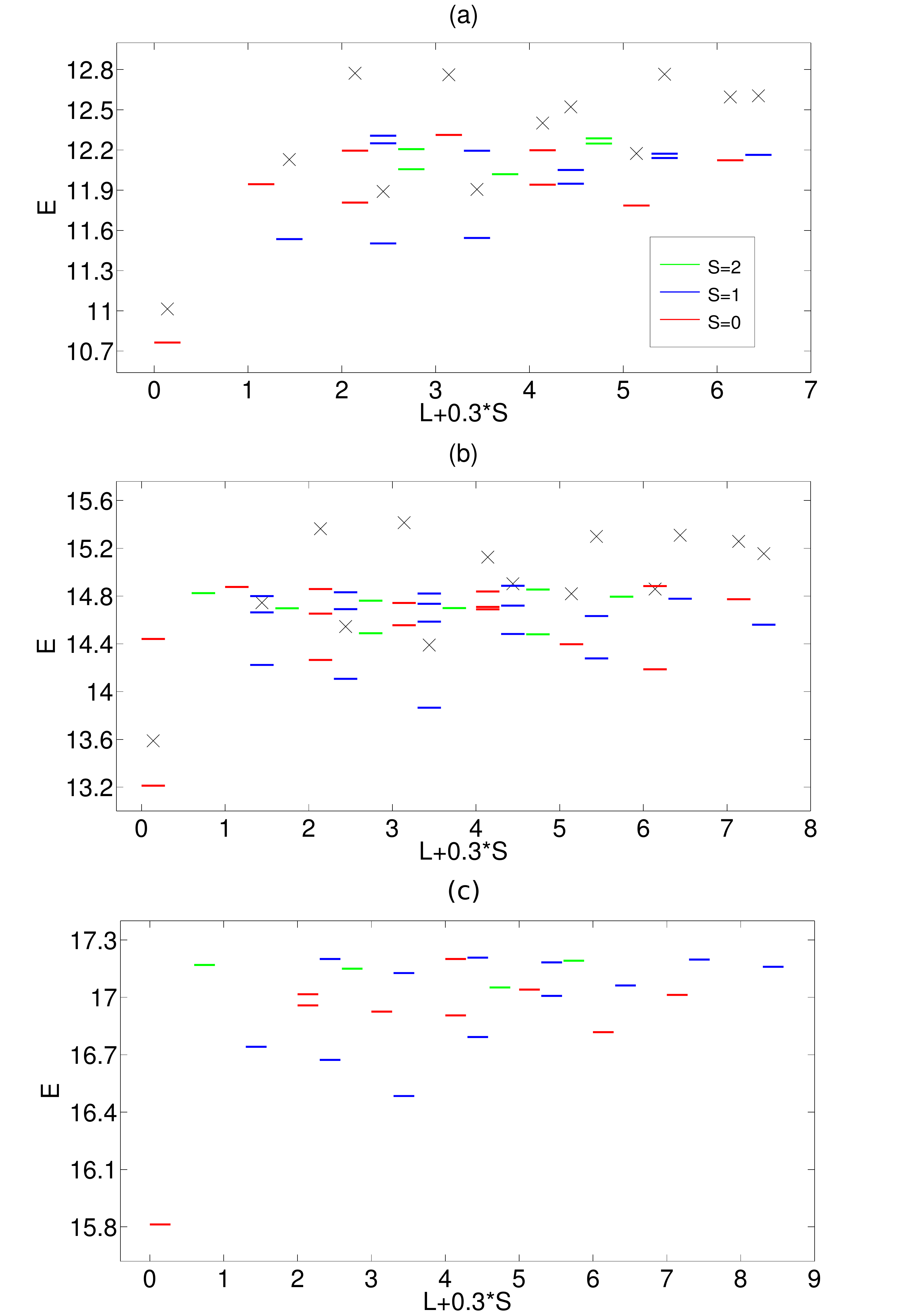}
\caption{Energy spectra of the $\nu=2$ ground states for the 2-body Hamiltonian $H^{\rm con}_2$. The cross represents the energy of the wave functions $\Psi^{[-1,-1]}_{2}$. (a) $N_{\uparrow}=6$, $N_{\downarrow}=6$ and $2Q=6$; (b) $N_{\uparrow}=7$, $N_{\downarrow}=7$ and $2Q=7$; (c) $N_{\uparrow}=8$, $N_{\downarrow}=8$ and $2Q=8$. The inset of panel (a) shows the color scheme for all panels.}
\label{Figure8}
\end{figure}

\begin{figure}
\includegraphics[width=0.45\textwidth]{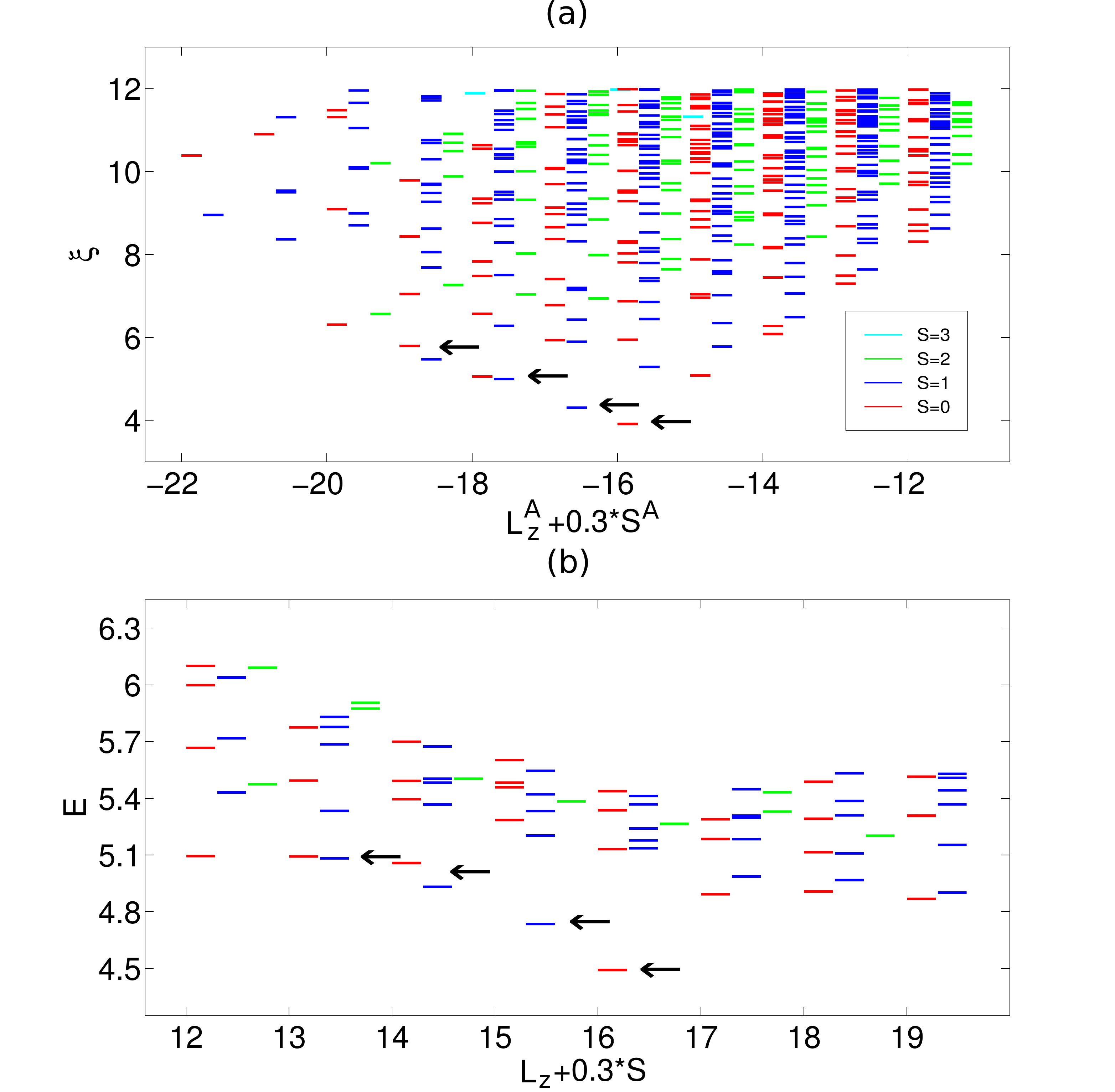}
\caption{RSES and edge excitations of the $\nu=2$ state. (a) RSES for the exact ground state of the 2-body Hamiltonian $H^{\rm con}_2$ for $N_{\uparrow}=8$, $N_{\downarrow}=8$, $N^A_{\uparrow}=4$, and $N^A_{\downarrow}=4$. (b) Energy spectrum on disk geometry of the Hamiltonian ${\widetilde H}_2$ for $N_{\uparrow}=4$ and $N_{\downarrow}=4$; the confinement potential parameter is taken to be $\omega_c=0.4$. The inset of panel (a) shows the color scheme for both panels. The arrows in panel (b) indicate the ground state and backward-moving edge modes and the arrows in panel (a) show the corresponding levels in the RSES.}
\label{Figure9}
\end{figure}

\begin{figure}
\includegraphics[width=0.45\textwidth]{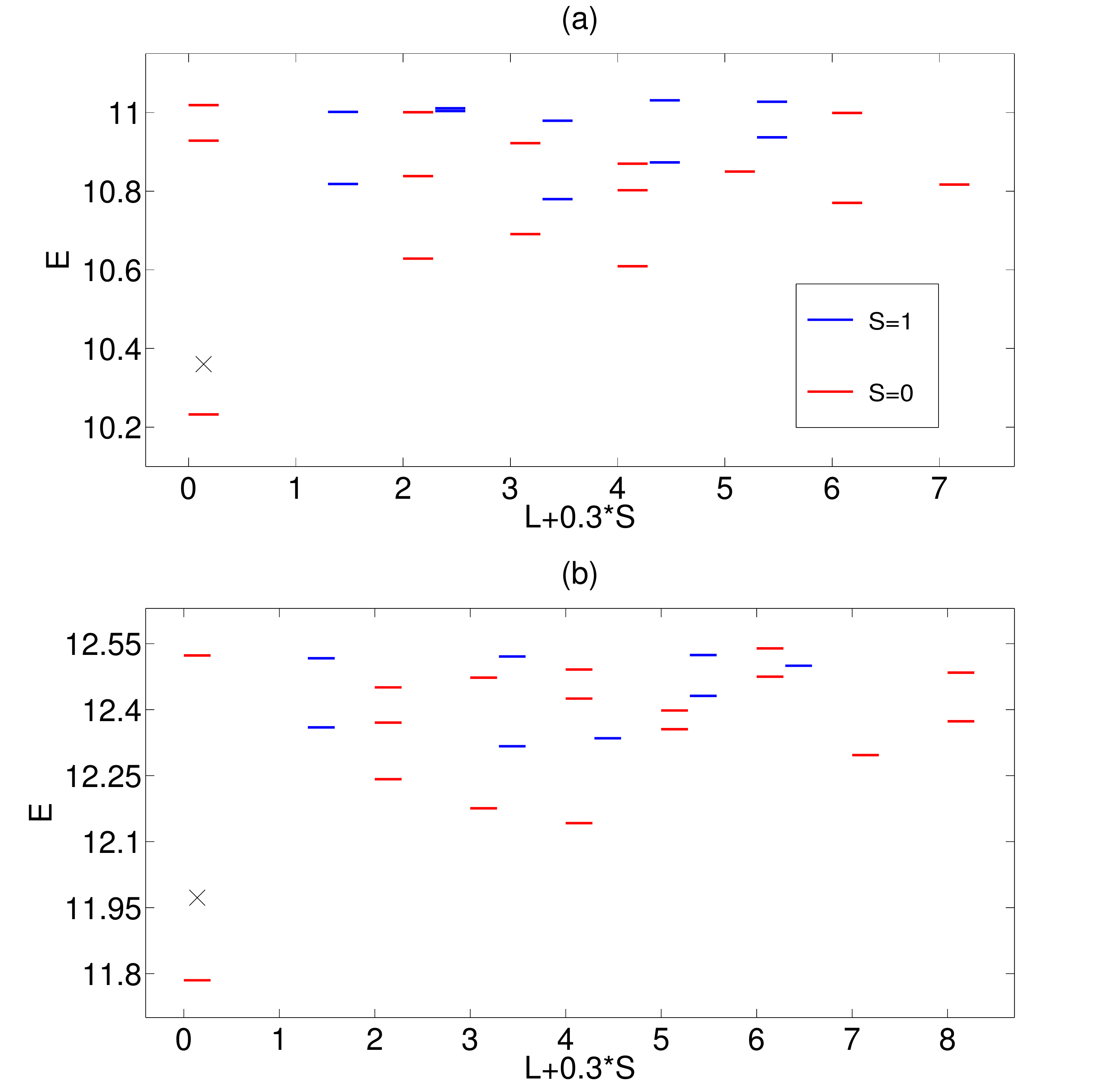}
\caption{Energy spectra of the $\nu=1$ ground states for the 2-body Hamiltonian $H_2$ with $c_0=1$, $c_2=0.3$ and all other $c_\alpha=0$ for $\alpha{\neq}0,2$. The crosses represent the energies of the wave functions $\Psi^{\rm JSS}_{1}$. (a) $N_{\uparrow}=6$, $N_{\downarrow}=6$ and $2Q=9$; (b) $N_{\uparrow}=7$, $N_{\downarrow}=7$ and $2Q=11$. The inset of panel (a) shows the color scheme for both panels.}
\label{Figure10}
\end{figure}

\begin{figure}
\includegraphics[width=0.45\textwidth]{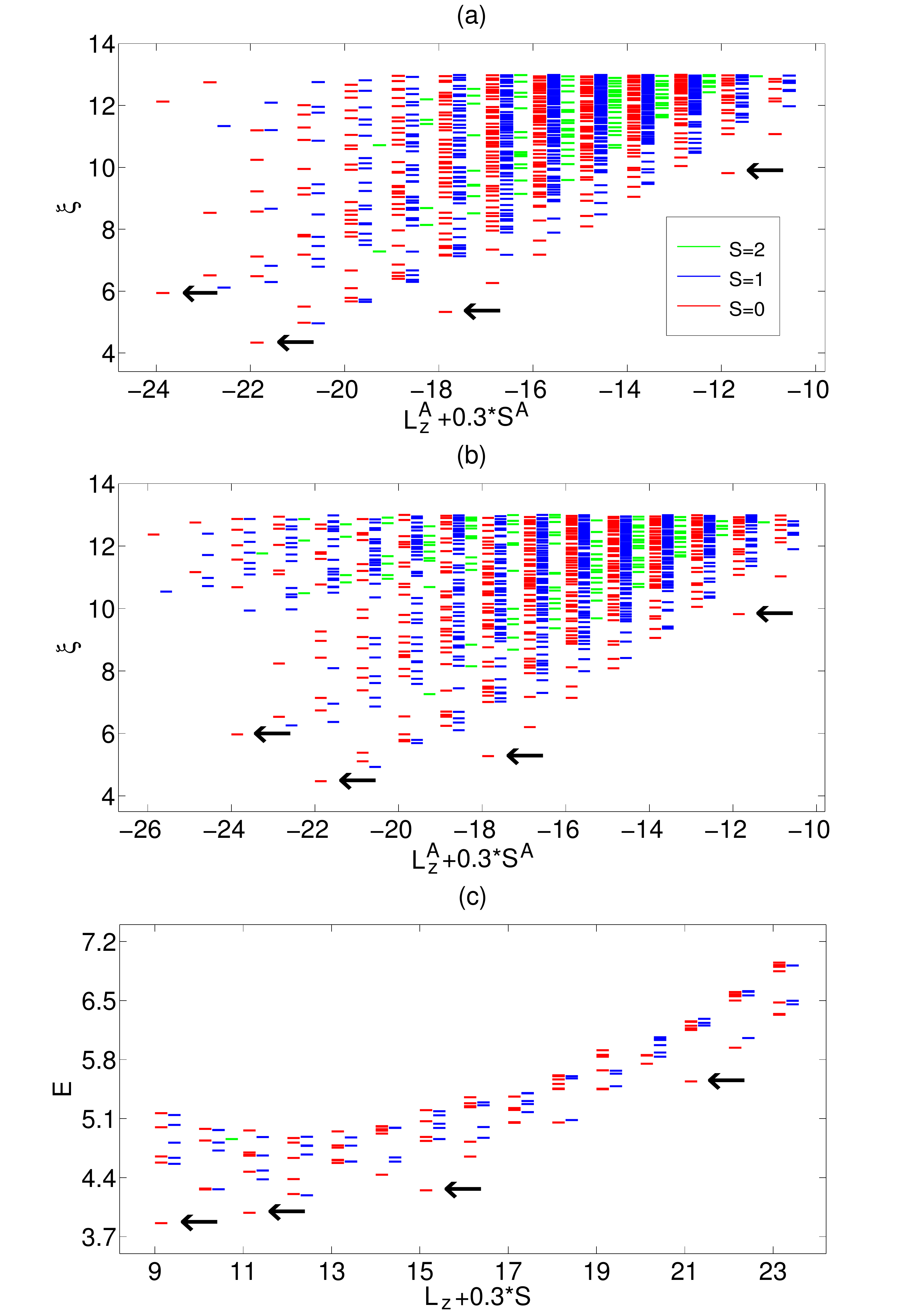}
\caption{RSES and edge excitations of the $\nu=1$ state. (a) RSES for $\Psi^{\rm JSS}_1$ for $N_{\uparrow}=7$, $N_{\downarrow}=7$, $N^A_{\uparrow}=3$, and $N^A_{\downarrow}=3$. (b) RSES for the ground state of the 2-body Hamiltonian $H_2$ for $N_{\uparrow}=7$, $N_{\downarrow}=7$, $N^A_{\uparrow}=3$, and $N^A_{\downarrow}=3$; the parameters of the Hamiltonian are $c_0=1$, $c_2=0.3$ and all other $c_\alpha=0$ for $\alpha{\neq}0,2$. (c) Energy spectrum on disk for $N_{\uparrow}=3$ and $N_{\downarrow}=3$ for the Hamiltonian ${\widetilde H}_2$ with ${\widetilde c}_2=0.3$ and the confinement potential parameter $\omega_c=0.4$. The inset of panel (a) shows the color scheme for all panels. The arrows in (c) indicate the states obtained with four different choices for $\Phi_2$ in the wave function Eq.~(\ref{JSSWave}), which are $[3,3]$, $[4,2]$, $[5,1]$ and $[6,0]$ (from left to right). The arrows in (a) and (b) show the corresponding levels in the RSES, which nicely match the starting points of various edge branches.}
\label{Figure11}
\end{figure}

\begin{table*}
\centering
\begin{tabular}{|c|cccccccccccc|}
\hline
       &       &       &       &       &       &$(L,S)$&       &       &       &       &       &       \\\hline
Figure & (0,0) & (0,1) & (1,0) & (1,1) & (2,0) & (2,1) & (3,0) & (3,1) & (4,0) & (4,1) & (5,0) & (5,1) \\\hline
 1(a)  & 1     & -     & -     & 0.994 & 0.994 & 0.998 & 0.997 & 0.997 & 0.997 & 0.997 & -     & -     \\
       & 49    &       &       & 203   & 161   & 302   & 180   & 438   & 261   & 518   &       &       \\\hline
 1(b)  & 1     & -     & -     & 0.993 & 0.992 & 0.997 & 0.996 & 0.997 & 0.997 & 0.997 & 0.996 & 0.996 \\
       & 713   &       &       & 4324  & 3122  & 6901  & 4099  & 9745  & 5375  & 12101 & 6216  & 14651 \\\hline
 1(c)  & -     & 0.992 & 0.992 & -     & -     & 0.988 & 0.983 & -     & 0.988 & 0.988 & -     & 0.990 \\
       &       & 969   & 1220  &       &       & 4476  & 2684  &       & 3234  & 7713  &       & 9026  \\\hline
 2(a)  & 0.997 & -     & -     & 0.973 & 0.974 & 0.953 & 0.980 & 0.972 & -     & -     & -     & -     \\
       & 16    &       &       & 53    & 41    & 70    & 39    & 107   &       &       &       &       \\\hline
 2(b)  & 0.992 & -     & -     & 0.984 & 0.987 & 0.968 & 0.947 & 0.971 & 0.983 & 0.983 & -     & -     \\
       & 2186  &       &       & 14764 & 10046 & 23908 & 13479 & 33359 & 17422 & 41880 &       &       \\\hline
 2(c)  & -     & 0.978 & 0.965 & -     & -     & 0.993 & 0.990 & -     & -     & -     & -     & -     \\
       &       & 363   & 447   &       &       & 1615  & 961   &       &       &       &       &       \\\hline
\end{tabular}
\caption{Overlaps between the trial states $\Psi^{[1,1]}_{2/3}$ and $\Psi^{[2,2]}_{4/5}$ and corresponding exact eigenstates shown in Figs.~\ref{Figure1} and~\ref{Figure2}. $L$ is the orbital angular momentum, $S$ is the spin quantum number, and ``$-$" means that there is no trial state in that $(L,S)$ sector. The total number of linearly independent $(L,S)$ multiplets is given below each overlap. The same conventions are used in all other tables.}
\label{Table1}
\end{table*}

\begin{table*}
\centering
\begin{tabular}{|c|cccccc|}
\hline
       &         &         & $(L,S)$  &          &         &         \\\hline
Figure & (0,2.5) & (1,2.5) & (2,2.5)* & (3,2.5)* & (4,2.5) & (5,2.5) \\\hline
 3(a)  & 0.995   & 0.984   & 1.390    & 1.371    & 0.954   & 0.977   \\
       & 1889    & 5628    & 9304     & 12857    & 16251   & 19432   \\\hline
\end{tabular}
\\
\begin{tabular}{|c|ccccccc|}
\hline
       &       &       &        &$(L,S)$&         &       &       \\\hline
Figure & (0,3) & (1,3) & (2,3)* & (3,3)* & (4,3)* & (5,3) & (6,3) \\\hline
 3(b)  & 0.794 & 0.856 & 0.783  & 0.468  & 0.801  & 0.731 & 0.767 \\
       & 5153  & 14812 & 24855  & 34029  & 43334  & 51546 & 59696 \\\hline
\end{tabular}
\caption{Overlaps between the trial states $\Psi^{[1,2]}_{3/4}$ and $\Psi^{[-1,-2]}_{3/2}$ and corresponding exact eigenstates shown in Fig.~\ref{Figure3}. The stars mark $(L,S)$ quantum numbers where the CF theory produces {\em two} independent states; the overlaps in these cases are defined as $\sqrt{\sum_{ij} \left[\langle\Psi^E_i|\Psi^T_j\rangle\right]^2}$ where the summation is over the lowest two exact states $|\Psi^E_i\rangle$ and trial states $|\Psi^T_j\rangle$ in the same $(L,S)$ sector. The total number of linearly independent $(L,S)$ multiplets is given below each overlap.}
\label{Table2}
\end{table*}

\begin{table*}
\centering
\begin{tabular}{|c|ccccccccccccc|}
\hline
       &       &       &       &       &       &       &$(L,S)$&       &       &       &       &       &       \\\hline
Figure & (0,0) & (0,1) & (1,0) & (1,1) & (2,0) & (2,1) & (3,0) & (3,1) & (4,0) & (4,1) & (5,0) & (6,1) & (7,0) \\\hline
 4(a)  & 0.985 & -     & -     & 0.949 & 0.933 & 0.933 & 0.971 & 0.964 & 0.918 & 0.947 & -     & -     & -     \\
       & 646   &       &       & 4117  & 2802  & 6619  & 3664  & 9258  & 4786  & 11494 &       &       &       \\\hline
 4(b)  & -     & 0.808 & 0.816 & -     & -     & 0.638 & 0.839 & -     & -     & 0.836 & 0.763 & 0.871 & 0.891 \\
       &       & 934   & 1064  &       &       & 4317  & 2326  &       &       & 7368  & 3407  & 9866  & 4235  \\\hline
 4(c)  & 0.965 & -     & -     & 0.943 & 0.914 & -     & -     & 0.934 & 0.923 & -     & -     & -     & -     \\
       & 4604  &       &       & 33132 & 21707 &       &       & 75440 & 37771 &       &       &       &       \\\hline
\end{tabular}
\caption{Overlaps between the trial states $\Psi^{[-2,-2]}_{4/3}$ and corresponding exact eigenstates shown in Fig.~\ref{Figure4}. $L$ is the orbital angular momentum, $S$ is the spin quantum number, and ``$-$" means that there is no trial state in that $(L,S)$ sector. The total number of linearly independent $(L,S)$ multiplets is given below each overlap.}
\label{Table3}
\end{table*}

\begin{table*}
\centering
\begin{tabular}{|c|cccccccccc|}
\hline
       &       &       &       &       &$(L,S)$&       &       &        &       &        \\\hline
Figure & (0,0) & (0,1) & (1,0) & (1,1) & (2,0) & (2,1) & (3,0) & (3,1)  & (4,0) & (4,1)  \\\hline
 5(a)  & 0.918 & -     & -     & 0.590 & 0.701 & 0.927 & 0.928 & 0.835  & -     & -      \\
       & 79    &       &       & 412   & 277   & 619   & 327   & 888    & -     & -      \\\hline
 5(b)  & -     & 0.696 & 0.871 & -     & -     & 0.737 & 0.647 & -      & -     & 0.417  \\
       &       & 934   & 1064  &       &       & 4317  & 2326  &        &       & 7368   \\\hline
 5(c)  & 0.897 & -     & -     & 0.534 & 0.610 & 0.738 & 0.727 & 0.782  & 0.622 & 0.795  \\
       & 6708  &       &       & 50057 & 31815 & 82111 & 43273 & 114205 & 55460 & 143987 \\\hline
 6(a)  & 1     & -     & -     & 0.936 & 0.969 & 0.995 & 0.993 & 0.992  & -     & -      \\
       & 79    &       &       & 412   & 277   & 619   & 327   & 888    & -     & -      \\\hline
 6(b)  & -     & 0.988 & 0.989 & -     & -     & 0.970 & 0.956 & -      & -     & 0.457  \\
       &       & 934   & 1064  &       &       & 4317  & 2326  &        &       & 7368   \\\hline
\end{tabular}
\caption{Overlaps between the NASS trial states $\Psi^{\rm NASS}_{4/3}$ (with excitations created within a spinful bipartite CF representation) and corresponding exact eigenstates shown in Figs.~\ref{Figure5} and~\ref{Figure6}. $L$ is the orbital angular momentum, $S$ is the spin quantum number, and ``$-$" means that there is no trial state in that $(L,S)$ sector. The total number of linearly independent $(L,S)$ multiplets is given below each overlap.}
\label{Table4}
\end{table*}

\begin{table}
\centering
\begin{tabular}{|c|cccccc|}
\hline
      &       &       & $c_1$ &       &       &       \\\hline
$c_2$ & 0.0   &  0.1  & 0.2   &  0.3  & 0.4   & 0.5   \\\hline
0.0   & 0.985 & 0.977 & 0.939 & 0.000 & 0.000 & 0.000 \\
      & 0.918 & 0.916 & 0.898 & 0.000 & 0.000 & 0.000 \\\hline
0.1   & 0.980 & 0.981 & 0.972 & 0.917 & 0.000 & 0.000 \\
      & 0.926 & 0.939 & 0.948 & 0.944 & 0.873 & 0.000 \\\hline
0.2   & 0.956 & 0.961 & 0.960 & 0.938 & 0.723 & 0.000 \\
      & 0.912 & 0.932 & 0.955 & 0.976 & 0.964 & 0.157 \\\hline
0.3   & 0.908 & 0.907 & 0.898 & 0.860 & 0.120 & 0.004 \\
      & 0.871 & 0.889 & 0.916 & 0.931 & 0.811 & 0.124 \\\hline
0.4   & 0.793 & 0.737 & 0.567 & 0.247 & 0.033 & 0.008 \\
      & 0.796 & 0.794 & 0.774 & 0.650 & 0.129 & 0.048 \\\hline
0.5   & 0.491 & 0.339 & 0.204 & 0.107 & 0.027 & 0.000 \\
      & 0.685 & 0.638 & 0.527 & 0.303 & 0.107 & 0.029 \\\hline
\end{tabular}
\caption{Comparing the JCF and NASS trial states at 4/3 ($\Psi^{[-2,-2]}_{4/3}$ and $\Psi^{\rm NASS}_{4/3}$, respectively) with the exact ground states at the corresponding flux ($2Q$) values as a function of interaction. The calculations are for $N_{\uparrow}=6$ and $N_{\downarrow}=6$ with respect to $c_1$ (columns) and $c_2$ (rows); we set $c_0=1$. The upper number in each block gives the overlap of $\Psi^{[-2,-2]}_{4/3}$ with the corresponding exact ground state. The lower number in each block gives the overlap of $\Psi^{\rm NASS}_{4/3}$ with the corresponding exact ground state.}
\label{Table5}
\end{table}

\begin{table*}
\centering
\begin{tabular}{|c|cccccccccccccc|}
\hline
       &       &       &       &       &       &       &$(L,S)$&       &       &        &       &       &       &      \\\hline
Figure & (0,0) & (1,1) & (2,0) & (2,1) & (3,0) & (3,1) & (4,0) & (4,1) & (5,0) & (5,1) & (6,0) & (6,1) & (7,0) & (7,1) \\\hline
 8(a)  & 0.943 & 0.765 & 0.503 & 0.847 & 0.838 & 0.881 & 0.683 & 0.682 & 0.902 & 0.720 & 0.754 & 0.834 & -     & -     \\
       & 36    & 163   & 111   & 240   & 122   & 345   & 175   & 401   & 173   & 479   & 216   & 507   &       &       \\\hline
 8(b)  & 0.888 & 0.812 & 0.517 & 0.867 & 0.622 & 0.808 & 0.433 & 0.866 & 0.862 & 0.555 & 0.761 & 0.833 & 0.815 & 0.771 \\
       & 164   & 989   & 639   & 1526  & 791   & 2169  & 1061  & 2620  & 1165  & 3149  & 1386  & 3471  & 1435  & 3850  \\\hline       
\end{tabular}
\caption{Overlaps between the trial states $\Psi^{[-1,-1]}_{2}$ and corresponding exact eigenstates shown in Fig.~\ref{Figure8}. $L$ is the orbital angular momentum, $S$ is the spin quantum number, and ``$-$" means that there is no trial state in that $(L,S)$ sector. The total number of linearly independent $(L,S)$ multiplets is given below each overlap.}
\label{Table6}
\end{table*}

\begin{table}
\centering
\begin{tabular}{|c|cccccc|}
\hline
      &       &       & $c_1$ &       &       &       \\\hline
$c_2$ & 0.0   &  0.1  & 0.2   &  0.3  & 0.4   & 0.5   \\\hline
0.0   & 0.888 & 0.915 & 0.939 & 0.950 & 0.000 & 0.000 \\
      & 0.161 & 0.168 & 0.069 & 0.002 & 0.000 & 0.000 \\\hline
0.1   & 0.843 & 0.877 & 0.914 & 0.948 & 0.949 & 0.000 \\
      & 0.825 & 0.750 & 0.640 & 0.387 & 0.053 & 0.000 \\\hline
0.2   & 0.767 & 0.000 & 0.849 & 0.908 & 0.952 & 0.007 \\
      & 0.916 & 0.897 & 0.847 & 0.709 & 0.363 & 0.032 \\\hline
0.3   & 0.652 & 0.000 & 0.687 & 0.681 & 0.151 & 0.028 \\
      & 0.940 & 0.942 & 0.934 & 0.897 & 0.609 & 0.003 \\\hline
0.4   & 0.000 & 0.000 & 0.431 & 0.268 & 0.071 & 0.011 \\
      & 0.925 & 0.922 & 0.894 & 0.774 & 0.479 & 0.035 \\\hline
0.5   & 0.000 & 0.000 & 0.256 & 0.144 & 0.053 & 0.019 \\
      & 0.876 & 0.852 & 0.774 & 0.602 & 0.364 & 0.001 \\\hline
\end{tabular}
\caption{Comparison of $\Psi^{[-1,-1]}_{2}$ at $\nu=2$ and $\Psi^{\rm JSS}_{1}$ at $\nu=1$ for $N_\uparrow=N_\downarrow=7$ with the corresponding exact ground states as a function of pseudopotential parameters. We set $c_0=1$ and vary $c_1$ (columns) and $c_2$ (rows). The upper number in each block gives the overlap of $\Psi^{[-1,-1]}_{2}$ with the corresponding exact ground state and the lower number of $\Psi^{\rm JSS}_{1}$ with the corresponding exact ground state.}
\label{Table7}
\end{table}

\section{Bosons at Fractional Fillings}
\label{fractional}

\subsection{$\nu=2/3$ and $4/5$}

(i) The Halperin $221$ state is the unique exact zero energy state of the hard-core interaction $H^{\rm con}_2$ at flux $2Q=3N/2-2$. The quasihole states, obtained by adding flux, are also exact zero energy states of $H^{\rm con}_2$, whose counting can be predicted in several ways and the wave functions are also known exactly~\cite{Ardonne3,Estienne}. 

Exact solutions are not known for the neutral excitations and the quasiparticles, which do not have zero energy with respect to $H^{\rm con}_2$. For these we use the trial wave functions $\Psi^{[1,1]}_{2/3}={\cal P}_{\rm LLL} [ \Phi_1(\{z^\uparrow\}) \Phi_1(\{z^\downarrow\}) J(\{z\}) ]$. The lowest energy neutral excitations correspond to a particle-hole excitation in one of the $\Phi_1$ factors. When the flux is reduced by one unit, each $\Phi_1$ factor on the right hand side contains one particle in the second LL. We construct $L$ and $S$ eigenstates by taking appropriate linear combinations. Fig.~\ref{Figure1} gives the energies (shown by crosses) of the trial wave functions of the neutral excitations in panels (a) and (b) and of quasiparticle excitations in panel (c). The overlaps between the trial wave functions and the exact eigenstates are shown in Table~\ref{Table1}. These comparisons show that the CF theory provides an excellent description of the excitations of the $2/3$ spin-singlet state.

(ii) The incompressible $\Psi^{[2,2]}_{4/5}$ state occurs at $2Q=5N/4-3$. We find that the system at this flux value is incompressible for up to $12$ particles as shown in panels (a) and (b) of Fig.~\ref{Figure2}. We have explicitly constructed the wave function $\Psi^{[2,2]}_{4/5}={\cal P}_{\rm LLL} [ \Phi_2(\{z^\uparrow_i\}) \Phi_2(\{z^\downarrow_i\}) J(\{z\}) ]$ for the ground states and excitations. Their energies are shown by crosses in Fig.~\ref{Figure2}, and their overlaps with the corresponding exact states are shown in Table~\ref{Table1}, which have excellent agreement. We note in passing that another candidate at $\nu=4/5$ is a spin-singlet Gaffnian state~\cite{Davenport3}, but it is likely to describe a gapless or critical state rather than an incompressible state since it is given by the conformal blocks of a non-unitary conformal field theory. 

For the $2/3$ state, the edge energy spectrum is trivial and it has been found that the counting of levels in RSES matches predictions~\cite{Ardonne3,Estienne}. In contrast, the edge spectrum of the $4/5$ state is expected to be complicated, containing several branches, because composite fermions occupy two $\Lambda$ levels. The studies of fermionic $2/5$ state tell us that such structures can only be seen for a rather large number of particles~\cite{Sreejith3,Rodriguez2}. The systems studied here are too small to bring out the edge physics.

\subsection{$\nu=3/4$ and $3/2$}

The $3/4$ state $\Psi^{[1,2]}_{3/4}$ occurs at $2Q=4N/3-8/3$ and the $3/2$ state $\Psi^{[-1,-2]}_{3/2}$ occurs at $2Q=2N/3+2/3$. These are spin-partially-polarized states. They are both derived from the spin-partially-polarized IQH state at $\nu^*=3$, one with parallel flux attachment and the other with reverse flux attachment. Fig.~\ref{Figure3} shows the energy spectra for the contact interaction $H^{\rm con}_2$ at these two filling factors and their comparison with the trial wave functions for the ground state as well as neutral excitations. Table~\ref{Table2} gives the overlaps of the trial states and exact states shown in Fig.~\ref{Figure3}. In some orbital and spin angular momentum sectors, there are two trial states and we define the overlap as $\sqrt{\sum_{ij} \left[\langle\Psi^E_i|\Psi^T_j\rangle\right]^2}$ where the summation is over the lowest two exact states $|\Psi^E_i\rangle$ and trial states $|\Psi^T_j\rangle$. These results show that the actual $3/4$ state is very well described by the CF theory, whereas this theory is less accurate for $3/2$. 

Note that the number of particles in each spin component is fixed (because the Hamiltonian $H^{\rm con}_2$ conserves the $z$-component of spin), so only states with total spin $S{\geq}|N_{\uparrow}-N_{\downarrow}|/2$ may occur. (Should we allow the spins to flip, these spin-partially-polarized state will not be ground states.) It is interesting to note that the low energy part of the spectrum contains states with $S=|N_{\uparrow}-N_{\downarrow}|/2$, with the states with higher values of $S$ appearing at much higher energies. This feature is nicely explained by the CF theory as follows. The $3/4$ and $3/2$ states map into $[1,2]$ and $[-1,-2]$ of composite fermions, and the lowest energy excitations (without changing $S_z$) contain a single CF exciton either in the spin-up sector or in spin-down sector. The resulting states satisfy the Fock condition (all occupied states in the spin-up sector are definitely occupied in the spin-down sector, and therefore the wave function is annihilated upon further antisymmetrization), and thus represent states with $S=|S_z|=|N_{\uparrow}-N_{\downarrow}|/2$. To produce a state with $S>|N_{\uparrow}-N_{\downarrow}|/2$ one must consider CF configurations containing at least two CF excitons, which are expected to lie at higher energies.

\subsection{$\nu=4/3$}

The filling factor $4/3$ has been considered~\cite{Grab,Furukawa} because it may provide a realization of the simplest NASS state $\Psi^{\rm NASS}_{4/3}$. At the same time, the CF theory provides another candidate $\Psi^{[-2,-2]}_{4/3}$ here. It is therefore of interest to ask what kinds of interaction would favor these states. The states $\Psi^{[-2,-2]}_{4/3}$ and $\Psi^{\rm NASS}_{4/3}$ occur at different shifts with $2Q=3N/4+1$ and $2Q=3N/4-2$, respectively, on the spherical geometry. 

Let us first consider the 2-body interaction. For the contact interaction $H^{\rm con}_2$, the spectrum for 12 particles at $2Q=3N/4+1$ is shown  in Fig.~\ref{Figure4}(a), and for $2Q=3N/4-2$ in Fig.~\ref{Figure6}(a). The overlaps of trial states and exact eigenstates are shown in Table~\ref{Table3} and~\ref{Table4}, respectively. Given that the JCF ground state has a higher overlap (0.985) than the NASS ground state (0.918) in spite of a larger Hilbert space (646 independent $L=S=0$ multiplets as opposed to 79 for NASS), these comparisons suggest that the states $\Psi^{[-2,-2]}_{4/3}$ is favored for the contact interaction. Comparison is also shown for excitations. 

To test the stability of the JCF and NASS states at filling factor $\nu=4/3$, we further test their performances when changing the coefficients $c_\alpha$ for $\alpha=1,2$ in the Hamiltonian $H_2$. The results are shown in Tables~\ref{Table5}. Both states remain good approximations for small values of $c_1$ and $c_2$, but are destroyed at large enough values for these parameters. We should emphasize that these numbers are not to be compared directly since the two states occur at different shift, and the dimensions of the subspaces with fixed $L$ and $S$ quantum numbers are different. 

As mentioned previously, the NASS is the exact ground state for the 3-body contact interaction $H_3$. The energy spectra corresponding to the NASS shift are shown in Fig.~\ref{Figure6} for this 3-body interaction. From the energy comparisons shown in this figure, and the overlaps shown in Table~\ref{Table4}, the excitations are very well described by the trial wave functions which create CF excitations in individual factors of Eq.~(\ref{NASSWave2}). 

We have also compared the RSES of the exact NASS state and the 2-body ground state in Fig.~\ref{Figure7}. The RSES are similar, as would be expected from the reasonably high overlaps. We also show the energy spectrum in the disk geometry, which, however, does not has very similar structure as the RSES. In fact, the energy spectrum in Fig.~\ref{Figure7}(c) is better understood as reverse-flux-attached CF state, as described below in Sec.~\ref{nu2}. We have not studied the RSES for $\Psi^{[-2,-2]}_{4/3}$ or the corresponding exact ground state. Since composite fermions occupy two $\Lambda$ levels in both spin sectors in the $4/3$ state, we do not expect the RSES to give very useful information using the system sizes that are accessible to exact diagonalization or for which $\Psi^{[-2,-2]}_{4/3}$ can be explicitly generated. 

We note that any spectrum in Figs.~\ref{Figure4} and~\ref{Figure5} can be interpreted in two different ways. For example, the $N_{\uparrow}=6$, $N_{\downarrow}=6$ and $2Q=7$ state in Fig.~\ref{Figure4}(a) can be thought of as excitations of $\Psi^{[-2,-2]}_{4/3}$, but here the NASS gives a satisfactory account of the exact spectrum. On the other hand, for Figs.~\ref{Figure4}(b) and~\ref{Figure5}(b), both interpretations work comparably well (although they predict different numbers of states), as seen from the overlaps in Tables \ref{Table3} and \ref{Table4}. 

Taking all of these results into account, while our studies do not rule out the NASS state, they suggest that the 4/3 ground state for the contact interaction is likely to be $\Psi^{[-2,-2]}_{4/3}$ with Abelian excitations. 

It would be useful to compare these two candidate states in the torus geometry where they compete directly. Recently, composite fermion wave functions have been successfully constructed in the torus geometry~\cite{Hermanns} for spin-polarized state at filling factors $2/3$ (for 10 bosons) and $2/5$ (for 6 fermions). Generalizing this method to spinful cases could be very interesting, although the numerical implementation of such schemes is expected to be very difficult. 

\section{Bosons at Integral Fillings}
\label{integral}

\subsection{$\nu=2$ state}
\label{nu2}

We consider the state $\Psi^{[-1,-1]}_{2}={\cal P}_{\rm LLL} [ \Phi_{-1}(\{z^\uparrow\}) \Phi_{-1}(\{z^\downarrow\}) J(\{z\}) ] $, obtained from the $\nu^*=2$ spin-singlet state with reverse flux attachment. This state is analogous to the spin-singlet 2/3 state of fermions~\cite{Wu93}. It has attracted special interest recently as an example of symmetry protected bosonic integer topological states~\cite{Chen,Lu,Senthil}, which refer to states with no topological order ({\em i.e.}, Abelian or non-Abelian fractional excitations) but are still topologically non-trivial.

As an initial test, we find that the ground state of the 2-body contact interaction $H^{\rm con}_2$ at the $2Q$ values corresponding to $\Psi^{[-1,-1]}_{2}$ indeed has $L=0$ and $S=0$ for up to $18$ particles. Fig.~\ref{Figure8} shows the energy spectra. [For 18 particles the dimension of the Fock space is very large (with 58,130,756 states in the $L_z=S_z=0$ sector), and producing eigenstates by the Lanczos method is computationally time consuming; we have only obtained the lowest few eigenstates to confirm that the ground state has $L=S=0$ and is separated from the excitations by a reasonable gap.] The overlaps of trial states and exact eigenstates for $N_{\uparrow}=N_{\downarrow}=6$ and $N_{\uparrow}=N_{\downarrow}=7$ are shown in Table~\ref{Table6}. For $12$ ($14$) particles, the exact ground state has overlap 0.943 (0.888) with $\Psi^{[-1,-1]}_{2}$; for $16$ particles we are not able to generate the trial state as explained in Sec.~\ref{model} B. We also study the stability of the state under addition of longer-range interaction. Table~\ref{Table7} shows the evolution of overlaps between $\Psi^{[-1,-1]}_{2}$ and exact ground state for a range of values of $c_1,c_2$ (with $c_0=1$), demonstrating that $\Psi^{[-1,-1]}_{2}$ remains a good description of the ground state for a wide range of parameters. We should point out that the trial wave functions for excitations are not as accurate as the ground states as one can see from Fig.~\ref{Figure8} and Table~\ref{Table6}.

We also study the RSES and the edge spectrum. The CF theory implies a behavior similar to that of the $\nu=2/3$ spin-singlet fermionic state, which has been studied in Refs.~[\onlinecite{Moore}] and~[\onlinecite{Mine}]. In particular, one expects a backward-moving mode that carries spin but no charge, and a forward-moving mode that carries charge but no spin~\cite{Moore,Mine}. We show in Fig.~\ref{Figure9} the RSES of the ground state of a bosonic system at $\nu=2$ with $N_{\uparrow}=N_{\downarrow}=8$ particles and the edge excitations of a system with $N_{\uparrow}=N_{\downarrow}=4$ particles on a disk. (We add a parabolic confinement potential of an appropriate strength in the disk geometry, to ensure that the ground state has the angular momentum given by $\Psi^{[-1,-1]}_{2}$.) We see a strong similarity between the RSES and the spectrum of edge excitations on the disk. In particular, the RSES nicely captures the backward-moving mode marked by the arrows in Fig.~\ref{Figure9}. The counting of states for the backward-moving modes is also consistent with that found for the $\nu=2/3$ spin-unpolarized fermionic states~\cite{Moore,Mine}. The forward-moving mode is not clearly idenfiable in both the RSES and the disk edge spectrum, as was also the case for spin-singlet $2/3$ state~\cite{Moore,Mine}; this can be understood by noting that the velocity of this mode is sufficiently large that it rapidly merges into the continuum for the small systems accessible to our study. 

The incompressibility at $\nu=2$ for bosons occurs because of interactions between them, and is therefore closer to the FQH (rather than the IQH) of fermions. One may ask what is the charge of the excitations. Identifying an isolated CF particle or CF hole in one of the factors of $\Phi_1$, it is straightforward to see that the charge excess associated with it is equal to a unit charge. 

\subsection{$\nu=1$ state}

We now explore the validity of $\Psi^{\rm JSS}_1 (\{z\})$. This state is an excellent description of the ground state at filling factor $\nu=1$ with $2Q=N-4$ if some amount of $c_2$ interaction is turned on, as shown Fig.~\ref{Figure10}. The evolution of overlap between trial states and exact ground states with the coefficients $c_1$ and $c_2$ of the Hamiltonian $H_2$ is shown in Table~\ref{Table7}. 

It is natural to construct wave functions for the excitations of $\Psi^{\rm JSS}_1 (\{z\}) = {\cal P}_{\rm LLL} [ \Phi_2(\{z\}) \prod_{i<j}(z^\uparrow_i-z^\uparrow_j) \prod_{i<j} (z^\downarrow_i-z^\downarrow_j) ]$ by analogy to excitations of either the factor $\Phi_2$ or one of the two Jastrow factors on the right hand side. We have constructed such wave functions for the excited states, but neither of them gives very accurate description of the excitations. 

We also study the RSES at $\nu=1$. Fig.~\ref{Figure11} shows the RSES of the JSS wave function and the exact ground state wave function for a certain choice of parameters ($c_0=1,c_2=0.3$ and all other $c$'s are set to zero) for 14 particles. The two have similar low-lying levels. For many trial wave functions that are exact zero energy solutions of certain simple pseudopotential Hamiltonians, such as the Laughlin or Moore-Read wave functions, the entanglement spectrum contains only universal levels, {\em i.e.}, all levels represent edge excitations. That, however, is not true in general. For electronic systems, the RSES of the exact Coulomb eigenstates at $1/3$ or $5/2$ contain ``non-universal" levels, as is also true of the either the exact states at $n/(2n+1)$ or the JCF wave functions for those states. The trial state $\Psi^{\rm JSS}_1 (\{z\})$ also has many non-universal levels as it is not the exact zero energy state of a simple pseudopotential Hamiltonian and its construction requires LLL projection. The RSES of the exact state contains even more ``non-universal" levels. We also show the edge excitation spectrum on disk geometry in Fig.~\ref{Figure11}, and some similarities between the RSES and edge spectrum can be seen even for such a small system. A noteworthy feature is that there are several branches of edge excitations, and the starting points of these branches (indicated by arrows in Fig.~\ref{Figure11}) match nicely in both the RSES and the edge spectrum. [Note that the minimum value of angular momentum in (a) and (b) is $-33$ while the minimum value in (c) is $0$, so the positions of the arrows in (a) and (b) match exactly with those in (c) if the angular momentum values in (a) and (b) are relabeled by adding $33$.]  The starting points of edge excitations can be simply predicted using parton method: they correspond to different choices for the number of particles $[N_1,N_2]$ in the two $\Lambda$ levels in the $\Phi_2$ part of Eq.~(\ref{JSSWave}), given in the figure caption. (While the starting points of the edge branches are identifiable, they quickly spread and merge into the non-universal part, making an identification of the edge states difficult.) The existence of multiple branches in the edge excitation spectrum and the RSES have been observed before for spin-polarized fermionic $2/5$ state~\cite{Sreejith3,Rodriguez2}, which is also due to the appearance of a $\Phi_2$ factor in the trial wave functions. In short, the RSES and edge studies provide support to the identification of the exact state with $\Psi^{\rm JSS}_1 (\{z\})$, and, in particular, bring out features that can be understood by analogy to two filled $\Lambda$Ls of composite fermions.

\section{Conclusion}
\label{conclusion}

We have carried out an extensive study of quantum Hall effect for two-component bosons, studying a number of candidate states at fractional as well as integral fillings. Here is a summary of our findings:

(i) We have shown that for 2/3 and 4/5, the wave functions $\Psi^{[1,1]}_{2/3}$ and $\Psi^{[2,2]}_{4/5}$ provide an accurate representation of the spin singlet states of the contact interaction, for the ground state as well as excitations. 

(ii) We have also considered partially polarized states at 3/4 and 3/2. For the former the state $\Psi^{[2,1]}_{3/4}$ provides an accurate description of the ground state and excitations. For 3/2, $\Psi^{[-2,-1]}_{3/2}$ is not accurate.  

(iii) For $\nu=4/3$ we consider two candidates, $\Psi^{[-2,-2]}_{4/3}$ and $\Psi^{\rm NASS}_{4/3}$, which have Abelian and non-Abelian excitations, respectively. Previous works~\cite{Grab,Furukawa} suggested that the NASS state is realized at this filling factor. We find, from a direct comparison with the exact solution, that $\Psi^{[-2,-2]}_{4/3}$ is more likely for the 2-body contact interaction.

(iv) For $\nu=4/3$, the NASS state has been known to be the exact ground state for a 3-body interaction. We find that the exact excited states of this 3-body interaction correspond to CF excitations in the individual factors, confirming a spinful bipartite CF structure for this state. 

(v) For $\nu=2$,  $\Psi^{[-1,-1]}_{2}$ provides an accurate representation of the exact ground state for in certain parameter range of the 2-body Hamiltonian. The trial wave functions for the excitations are less accurate. The RSES and the disk energy spectrum provide a consistent description of the edge structure, both nicely displaying a backward-moving edge mode (which is similar to that found previously for the fermionic $2/3$ spin-singlet state~\cite{Moore,Mine}).

(vi) For $\nu=1$, the JSS state accurately represents the ground state for a 2-body interaction that contains terms beyond contact interaction. The RSES and edge excitation studies provide further confirmation of the validity of $\Psi_1^{\rm JSS}$, and, in particular, demonstrate the existene of several edge branches, which are fully consistent with the expectation from an underlying two-filled $\Lambda$L state. Our trial wave functions are not very accurate for the excitations. 

(vii) We note a systematic effect as a function of the filling factor: the agreement between the CF and the exact spectra becomes worse with increasing filling factor. There are two possible reasons for that. One, the JCF wave functions of the states with large fillings, e.g. $\nu=3/2, 4/3$ and 2, namely $\Psi^{[-2,-1]}_{3/2}$, $\Psi^{[-2,-2]}_{4/3}$ and $\Psi^{[-1,-1]}_{2}$, all require reverse flux attachment. We have found that for spinful particles, the wave functions involving reverse flux attachment are less accurate than those with parallel flux attachment. Two, from general arguments one expects that bosons at very high fillings are not in the FQH regime, because vortex lattice or other weakly interacting states may be preferred energetically. 
\\
\\
{\em Note added} --- At the time of the completion of this manuscript, we noticed the preprint~[\onlinecite{Furukawa2}], which also studies the $\nu=2$ state and has some overlap with Sec.~\ref{nu2} of our paper. Another preprint~[\onlinecite{Nicolas}] about the $\nu=2$ state has also appeared since then. These studies complement one another to an extent, as we briefly describe. All three works provide evidence for an incompressible state at filling factor $\nu=2$ for two-component bosons.  Refs.~[\onlinecite{Furukawa2}] and~[\onlinecite{Nicolas}] report finite-size scaling of gaps and consider the cases where the interaction is not $SU(2)$ invariant. Ref.~[\onlinecite{Furukawa2}] uses Chern-Simon field theory to interpret the counting on RSES; we compare the RSES with the edge spectrum on the disk geometry, and
also ﬁnd that edge excitations can be understood using CF theory. Ref.~[\onlinecite{Nicolas}] studies the ground state degeneracy on torus to rule out a competing non-Abelian state, whereas our observation that incompressible states occur for all cases with $N_\uparrow=N_\downarrow$ in the spherical geometry also rules out the non-Abelian state. We have also constructed and studied explicit trial wave functions for the ground states as well as excitations.

\section*{Acknowledgement}

We are immensely grateful to the authors, especially N. Regnault, of the DiagHam package for sharing their programs.  We thank Kai Sun for many useful discussions, the DOE for financial support under Grant No. DE-SC0005042, and Research Computing and Cyberinfrastructure, a unit of Information Technology Services at The Pennsylvania State University, for providing high-performance computing resources and services.

\end{document}